\documentclass[aps,prx,twocolumn,superscriptaddress,nofootinbib,showpacs]{revtex4-2}
\usepackage{amsmath}
\usepackage{amssymb}
\usepackage{graphicx}
\usepackage{xcolor,color}
\usepackage{url}
\usepackage{bm}
\usepackage{mathrsfs}
\usepackage[utf8]{inputenc}
\usepackage{hyperref}
\usepackage{enumerate}
\usepackage{amsthm}
\usepackage{bbm}
\usepackage[normalem]{ulem}
\usepackage{upgreek}
\usepackage{tensor}
\usepackage{siunitx}
\usepackage{orcidlink}
\usepackage[caption=false]{subfig}
\usepackage{float}
\usepackage{colortbl}

\definecolor{colour1}{HTML}{0571b0} 
\definecolor{colour2}{HTML}{92c5de} 
\definecolor{colour3}{HTML}{f4a582} 
\definecolor{colour4}{HTML}{ca0020} 
\definecolor{colour5}{HTML}{fe4a49} 
\definecolor{colour6}{HTML}{2d3092} 

\hypersetup{colorlinks=true, linkcolor=colour1, citecolor=colour1,
filecolor=colour1, urlcolor=colour1}

\theoremstyle{plain}

\usepackage{color,xcolor}
\usepackage{ulem}
\usepackage{float}
\usepackage{subfig}
\usepackage{cleveref}
\usepackage{multirow}
\usepackage{booktabs}
\usepackage{booktabs}
\usepackage{threeparttable}

\usepackage{txfonts}

\usepackage[export]{adjustbox}
\usepackage{times}
\usepackage{commath}

\crefname{equation}{Eq.}{Eq.}
\Crefname{equation}{Eqs.}{Eqs.}

\Crefname{section}{Sec.}{Sec.}

\Crefname{figure}{Fig.}{Fig.}

\usepackage{threeparttable}

\begin{document}

\title{Radiation fluxes of gravitational, electromagnetic, and scalar perturbations in type-D black holes: an exact approach}

\author{Changkai Chen\,\orcidlink{0000-0002-4023-0682},}
\affiliation{Department of Physics, Key Laboratory of Low Dimensional Quantum Structures and Quantum Control of Ministry of Education, Synergetic Innovation Center for Quantum Effects and Applications, Hunan Normal University, Changsha, 410081, Hunan, China}

\author{{Jiliang} {Jing}\,\orcidlink{0000-0002-2803-7900}, } 
\email[Corresponding author: ]{jljing@hunnu.edu.cn}
\affiliation{Department of Physics, Key Laboratory of Low Dimensional Quantum Structures and Quantum Control of Ministry of Education, Synergetic Innovation Center for Quantum Effects and Applications, Hunan Normal University, Changsha, 410081, Hunan, China}

\date{\today}
\begin{abstract}
We present a novel method that solves Teukolsky equations with the source to calculate radiation fluxes at infinity and event horizon for any perturbation fields of type-D black holes.
For the first time, we use the confluent Heun function to obtain the exact solutions of ingoing and outgoing waves for the Teukolsky equation.
This benefits from our derivation of the asymptotic analytic expression of the confluent Heun function at infinity.
It is interesting to note that these exact solutions are not subject to any constraints, such as low-frequency and weak-field.
To illustrate the correctness, we apply these exact solutions to calculate the gravitational, electromagnetic, and scalar radiations of the Schwarzschild black hole.
Numerical results show that the proposed exact solution appreciably improves the computational accuracy and efficiency compared with the 23rd post-Newtonian order expansion and the Mano-Suzuki-Takasugi method.

\end{abstract}
\maketitle

\section{Introduction}
Black hole (BH) perturbation theory  \cite{10.1143/PTPS.128.1,Sasaki_2003} is a method used to study various real relativistic objects, such as massive compact objects, jets, supernova explosions, binary systems, etc. Originally developed as a metric perturbation theory, for Schwarzschild BHs, Regge \cite{Regge_1957} and Zerilli \cite{Zerilli_1970} decoupled and separated a single master equation for the metric perturbation into odd and even parity parts, respectively.
However, no such equation has been established for rotating Kerr BHs.
This led Bardeen and Press \cite{Bardeen1973RadiationFI} to derive a master equation for the curvature perturbation of a Schwarzschild BH without the source ($T_{\ell m \omega}=0$) using the Newman-Penrose null-tetrad formalism, where the tetrad components of the curvature tensor serve as fundamental variables.
Expanding to a Kerr BH with the source ($T_{\ell m \omega}\neq 0$), Teukolsky \cite{Teukolsky1973} derived the curvature perturbation equation. This resulting equation, known as the Teukolsky equation, represents a wave equation for the null-tetrad component of the Weyl tensors $ \psi_0$ and $\psi_4$.
The Teukolsky equation describes the dynamics of various fields of  different spins as perturbations (scalar, neutrino, electromagnetic, and gravitational perturbations) to the Kerr metric. Therefore, the Teukolsky equation can be used as a mathematical model of gravitational waves (GWs) to construct GW waveform templates.
In recent years  there is an increased interest in GW detection, and we anticipate gaining a deeper understanding of BHs' demographics and properties in the coming years \cite{Abbott2016,Abbott2016a,Abbott2019,Abbott2021,Abbott2023}.
Future space-based GW detectors, such as the Laser Interferometer Space Antenna (LISA) \cite{Audley:2017drz}, TianQin \cite{Luo_2016,Mei_2020}, and Taiji \cite{Gong_2021,Ruan_2020}, will be built with the specific purpose of detecting GW signals from sources that radiate in the millihertz ($mHz$) bandwidth, namely extreme mass ratio inspirals (EMRIs) \cite{Amaro_Seoane_2018,AmaroSeoane2020,Isoyama_2022}.
BH perturbation theory has been utilized in modeling EMRI systems and their associated GWs \cite{Poisson:2011,Pound:2021}.
Under the influence of gravitational radiation reaction, the compact secondary body undergoes a slow inspiral into the primary supermassive black hole, emitting GWs that propagate to infinity.
Due to potential interference from other sources concurrently emitting GWs in the $mHz$ bandwidth \cite{Audley:2017drz}, detection and parameter estimation of EMRI signals will rely on matched filtering techniques.
Accurate calculation of GW waveform templates is thus of paramount importance, requiring precision on the order of fractions of radians in GW phases \cite{Babak_2017,Barack_2019}.
Upon precise modeling and interpretation, the obtained data presents an opportunity for unparalleled experiments examining general relativity and detecting hitherto unknown astrophysical phenomena \cite{Amaro-Seoane_2007,Fan:2020zhy,Zi:2021pdp}.

The Teukolsky equation without the source simplifies the perturbation problem significantly because it does not need to construct Green's function. This homogenous equation has been widely adopted to study the quasinormal modes (QNMs) of BHs and other relativistic celestial bodies, including massive dense stars, jets, supernova explosions, and more.
By exploiting infinite series of special functions, Leaver derived the analytical expressions for the solutions of the Regge-Wheeler (RW) and Teukolsky equations \cite{Leaver1986}, which enabled precise computation of the QNMs spectrum of black holes \cite{leaver1985a} and discussed their excitation using Green's function method \cite{Leaver_1986a}. But Green's function constructed by Leaver cannot be applied to calculate gravitational radiation.

It is known, but  historically somewhat under-appreciated in the physics literature, that both the RW and Teukolsky equations without the source are examples of the confluent Heun equation \cite{Ronveaux:1995,slavyanov2000special}.
One possible reason for the historical lack of attention given to the confluent Heun equation in physics literature is the difficulty in numerically calculating its associated Heun class functions in the past, leading to a prevalence of Leaver's solution.
However, recent advances in numerical algorithms have enabled the computation of Heun class functions using various mathematical software packages. Maple software version 7 introduced numerical calculations of Heun class functions in 2001.
And Motygin provided MATLAB codes for general and confluent Heun functions in 2015 \cite{Motygin_2015} and 2018 \cite{Motygin2018}, respectively.
The 12.1 version of Mathematica software released in 2020 also includes numerical calculation capabilities for Heun class functions. It should be noted, however, that the performance and accuracy of Heun class functions can differ across software packages.
Fiziev provided the analytical solutions \cite{Fiziev_2006,Fiziev_2007,Fiziev_2010,fiziev2009classes} to the source-free perturbation equations (RW or Teukolsky equations) in terms of the confluent Heun function.
Furthermore, these solutions have been utilized to calculate QNMs of the Teukolsky equation describing the Schwarzschild black hole\cite{Fiziev_2011a}, including its continuous spectrum\cite{borissov2010exact}, as well as for the central engine of Gamma-ray bursts (GRB) and cosmic jet of the Kerr black hole\cite{fiziev2009new,Staicova_2010}.
Following Fiziev's work, numerous researchers have employed Heun class functions to study source-free perturbation equations in various spacetimes, such as QNMs \cite{Semra2019} of Dirac field in $2+1$ dimensional GW background \cite{Zhang2014BlackHA}, QNMs and the reflection coefficient \cite{Horta_su_2021a} of massless fields in Kerr-Newman–de Sitter BH \cite{Hatsuda_2021}, and QNMs of the massive scalar field in Kerr-AdS$_5$ BH \cite{Noda2022}. Moreover, Cook et al. \cite{Cook_2014} converted the modes equations of Kerr BH into the confluent Heun equation and then solved the radial equation using the continued fraction method, and the angular equation using the spectral method, resulting in improved accuracy for the QNMs.
\begin{figure*}[htbp]
  \includegraphics[width=6.2in]{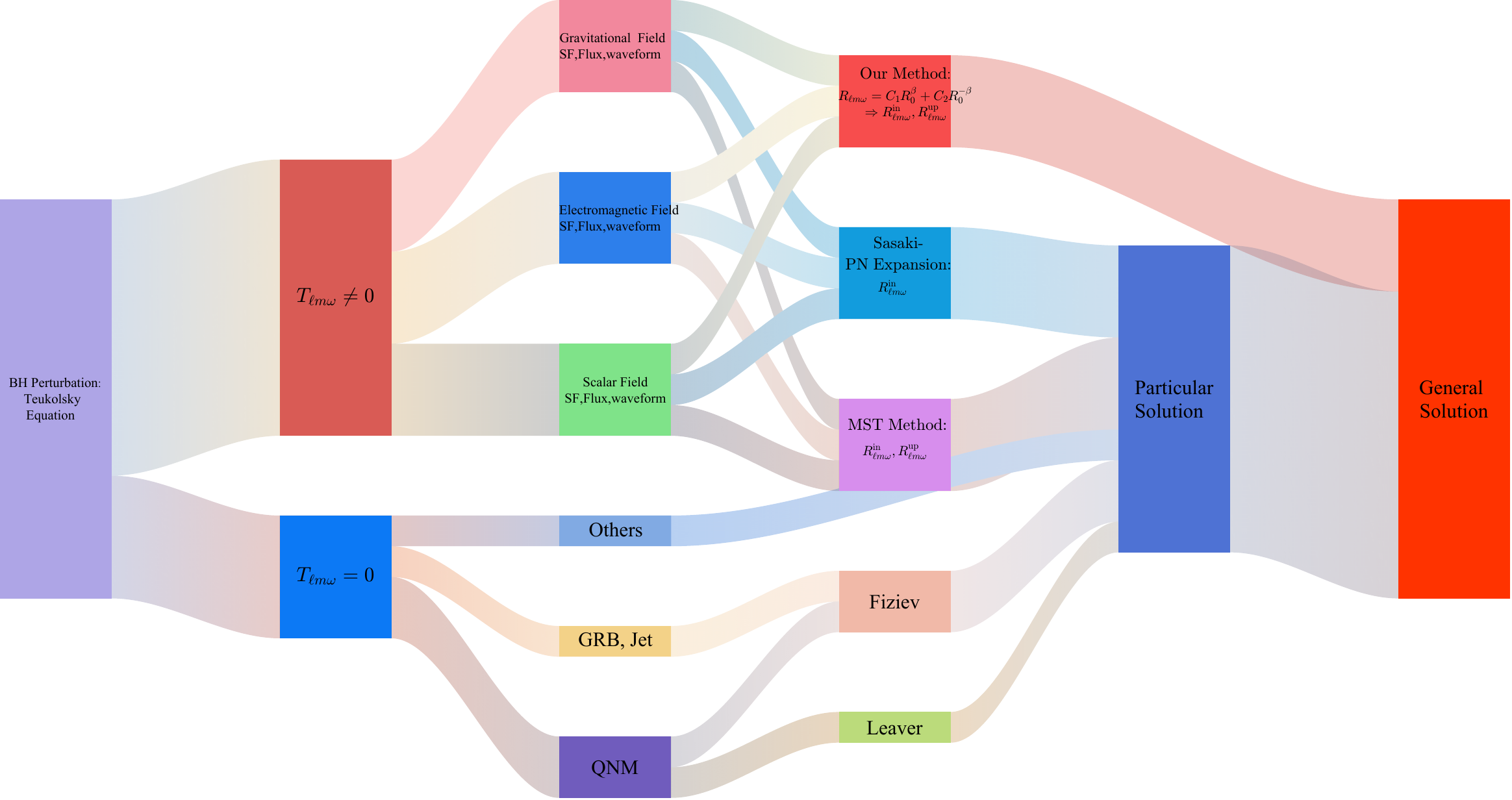}
  \caption{The difference between our method and other methods in black hole perturbation}\label{fig:introduction}
  \end{figure*}

In previous research, the construction of Green's function for the Teukolsky equation with the source was deemed analytically infeasible. As a result, numerical integration methods were widely employed by most scholars initially to investigate the radiation phenomenon associated with black hole perturbation.
These numerical methods involve simulating the propagation of GWs in a background spacetime distorted by a rotating BH with an external particle or other perturbation included.
Such simulations furnish invaluable insight into the behavior of gravitational waves and their interactions with black holes\cite{Press1973,Chrzanowski_1975,Tashiro_1981,chandrasekhar1998mathematical,Nakamura_1987}.
Subsequently, using the post-Newtonian (PN) expansion, Sasaki derived a part of Green's function \cite{Sasaki_1994} for the RW equation.
This derivation involved only the ingoing wave solution $X_{\ell m\omega }^{{\rm{in}}}$ and its asymptotic amplitudes at infinity.
Utilizing the Sasaki-Nakamura (SN) transformation, Sasaki transformed $X_{\ell m\omega }^{{\rm{in}}}$ into the ingoing wave solution $R_{\ell m\omega }^{{\rm{in}}}$ for the Teukolsky equation with the source in a Schwarzschild BH \cite{Tagoshi_1994} and a Kerr BH \cite{Shibata_1995}.
This approach is applied to determining $R^{\rm in}_{\ell m\omega}$ of a Schwarzschild BH up to 5.5PN order \cite{Tanaka_1996}, as well as $R^{\rm in}_{\ell m\omega}$ of a Kerr BH up to 4PN order \cite{10.1143/PTPS.128.1,Tagoshi:1996gh}.
For the absorption and tail correction of a Schwarzschild BH, it is necessary to construct another part of Green's function containing $R_{\ell m\omega }^{{\rm{up}}}$, which represents the homogeneous solution of the Teukolsky equation of pure outgoing waves.
Therefore, Poisson and Sasaki \cite{Poisson_1995} utilized the spherical Hankel function to construct only 1PN $X_{\ell m\omega }^{{\rm{up}}}$, but this solution is controversial and does not satisfy the conservation of the Wronskian for Kerr BHs \footnote{It follows from the conservation of the Wronskian that $X_{\ell m\omega }^{{\rm{up}}}(r \to 2M)\sim - \bar A_{\ell m\omega }^{{\rm{up }}}{e^{ - i\omega {r^*}}} + A_{\ell m\omega }^{{\rm{in }}}{e^{i\omega {r^*}}},A_{\ell m\omega }^{{\rm{in }}} \ne 0$. But $X_{\ell m\omega }^{{\rm{up}}}$ constructed using spherical Hankel functions has the amplitude $A_{\ell m\omega }^{{\rm{in }}} = 0$, which fails to satisfy this conservation relation.}.

The Mano-Suzuki-Takasugi (MST) method is a second analytical approach for constructing Green's function of the Teukolsky equation with the source.
Japanese researchers `revamped' Leaver's solutions and developed new series solutions for the radial solutions \cite{Mano1996RWE,Mano_1996,Sasaki_2003}.
The primary distinction between the MST solution and Leaver's solutions is that the former can obtain analytical expressions for asymptotic amplitudes, which the five solutions produced by Leaver do not satisfy.
The series solutions employed by the MST method, known as the MST expansions, are naturally adapted to carrying out low-frequency expansions. Casals believed that the MST series converge theoretically for any frequency value, albeit their convergence speed diminishes as the frequency magnitude increases \cite{Casals_2015}.
Initially, Mano et al. \cite{Mano1996RWE,Mano_1996} only presented the MST method as the second-order post-Minkowskian expansion result. Subsequently, Fujita numerically computed the renormalized angular momentum of the MST method and achieved remarkably precise solutions \cite{Fujita_2004,Fujita_2005}. Additionally, Fujita utilized the MST method for computing gravitational radiation at arbitrary high PN order.
For instance, Fujita provided 22PN results in a Schwarzschild BH \cite{Fujita_2012} and 11PN results in a Kerr BH\cite{Fujita_2015a}. Other than that, Fujita presented 5.5PN GW polarizations and associated factorized resumed waveforms\cite{Fujita_2010}. The MST method can also be employed to compute the self-force (SF) acting on point particles \cite{Poisson:2011,Fujita_2015GSF,Hikida_2004,Hikida_2005,Casals_2013}.

Apart from the MST method, there exist alternative approaches for computing the gravitational wave energy flux.
Fully relativistic GW fluxes from orbits of non-spinning particles were initially computed for eccentric orbits around a Schwarzschild BH \cite{Cutler:1994} and circular equatorial orbits around a Kerr BH \cite{Finn:2000}. Glampedakis et al. \cite{Glampedakis_2002} calculated Fluxes from eccentric orbits in the Kerr spacetime.
Fully generic GW fluxes from non-spinning particles were derived by Drasco\cite{Drasco_2006}, and Hughes et al. computed adiabatic waveforms of EMRIs \cite{Hughes:2021}.
Circular orbits in a BH spacetime with the spin of the secondary taken into account were investigated via GW fluxes in
Refs.~\cite{Han:2010,Harms:2016a,Harms:2016b,Lukes-Gerakopoulos:2017,Akcay:2020}, while quasi-circular adiabatic evolution of such orbits that includes spin was presented in Refs.~\cite{Piovano:2020,Piovano:2021,Skoupy:2021a,Rahman:2023}.
The first-order SF for circular orbits in the Schwarzschild spacetime was calculated in Ref.~\cite{Mathews:2022}.
Moreover, Skoupy et al. \cite{Skoupy:2021b} computed the fluxes from spinning bodies on eccentric equatorial orbits around a Kerr BH, and provided the linear in spin approximation for the adiabatic evolution \cite{Skoupy:2022}.
Referring to the frequency-domain method developed in Refs.~\cite{Drummond:2022a, Drummond:2022b} for generic orbits of spinning bodies around a Kerr black hole, Hughes et al. \cite{skoup2023asymptotic} computed the asymptotic GW fluxes from a spinning body moving on such orbits up to linear order in the secondary spin in a Kerr BH.

In this work, \textsl{we solve the Teukolsky equation with the source for the first time using the confluent Heun function to obtain an exact method for calculating gravitational, electromagnetic and scalar radiation fluxes for any type-D BHs.}
Our method can get the complete Green's functions including $R_{\ell m\omega }^{{\rm{in,up}}}$.
Different from the PN expansion and MST method in the BH perturbation approach, our strategy involves expressing the general solution of the homogeneous Teukolsky equation as a linear combination of two linearly independent particular solutions (constructed in the form of confluent Heun functions), and then using the boundary conditions to obtain $R_{\ell m\omega }^{{\rm{in,up}}}$.
The construction difference between our method and other methods is shown in \Cref{fig:introduction}.

The remainder of the paper is arranged as follows.
In \Cref{sec:GTF}, the Teukolsky equation is rewritten into a more general form, which can cover many curvature perturbation equations of Type-D spacetimes.
In \Cref{sec:Flux-GFM}, we revamped  Green's function method to obtain radiation fluxes for any perturbation fields.
For Teukolsky equations without the source, we proposed new analytical solution constructed by confluent Heun functions in \Cref{sec:GsolGFRTE}.
We present the analytical expression for the confluent Heun function at infinity for the first time in \Cref{sec:AsympForHeunC}.
To verify the correctness of our method, our method is applied to calculate the fluxes of scalar, electromagnetic, and gravitational perturbations for Schwarzschild BHs in \Cref{sec:Application}.
\Cref{sec:results} is devoted to assessing the accuracy and effectiveness of the method proposed in this work. To this end, we compare our results with those of post-Newtonian expansion and MST methods to validate the applicability of our approach.
The conclusions are presented in \Cref{sec:Conclusion}.
In this paper, we use geometrized units: $c=G=1$.


\section{Teukolsky Equations of Type-D BHs}\label{sec:GTF}
We now reformulate radial Teukolsky equations (RTE) into a more comprehensive form that can encompass all previously identified Teukolsky equations of Type-D BHs, which, names the general form of radial Teukolsky equation (GFRTE), is expressed as

\begin{equation}\label{eq:GFoTRE}
 \left[ {\Delta _n^{ - s + 1}\frac{d}{{dr}}\Delta _n^{s + 1}\frac{d}{{dr}} + V(r)} \right]{R_{\ell m\omega }} = {\Delta _n}{{}_{s}T_{\ell m\omega }},
\end{equation}
where
\begin{eqnarray}\label{eq:Delta0}
 && \Delta_n  = \sum\limits_{i = 0}^n  {{b_i}{r^{2 - i}}}  = \prod\limits_{i = 1}^n  {{{\left( {r - {r_i}} \right)}}},
\\ \label{eq:VV}
 && V(r) = \sum\limits_{i = 0}^\infty  {{{\bf v}_i}{r^i}}.
\end{eqnarray}

The Newman-Penrose formalism allows for the determination of the explicit form \eqref{eq:VV} of $V(r)$, and it can be observed that the simplest expansion of $V(r)$ does not contain any terms with fractional powers of $r$.
The common approach is to approximate $\Delta_n$ by $\Delta_2$ and $\Delta_4$ \footnote{The Teukolsky equation of Kerr-Newman (anti-)de Sitter BHs \cite{Khanal_1983,Suzuki_1998,Suzuki_1999} without the source corresponds to $\Delta_4$-type.}, resulting in various types of the GFRTEs with different forms in the potential term $V(r)$. This paper specifically focuses on the Teukolsky equation of $\Delta_2$-type, that is $\Delta_2  = \left( {r - {r_-}} \right)\left( {r - {r_+}} \right)$. Here,  $r_- $ is the inner horizon, and $r_+ $ is the outer (event) horizon.

For different black holes, the values of $\Delta_2$ and $V(r)$ in the Teukolsky equation will change. Now, we collect information on seven black holes, and the corresponding $\Delta_2$ and $V(r)$ are organized as follows:

 \textcolor{colour1}{\uppercase\expandafter{\romannumeral1}. Schwarzschild BHs}

The potential $V(r)$ of Schwarzschild BHs for all perturbation fields \cite{Bardeen1973RadiationFI,Tagoshi_1994} is
\begin{equation}\label{eq:ScharzchildV}
{V_{{\rm{Sch}}}} = {\omega ^2}{r^4} + 2is\omega {r^2}(r - 3M) - {\Delta _2}{\lambda},
\end{equation}
where ${r_-} = 0$, ${r_+}=r_{\rm H}=2M$ and  ${\lambda }$ is given in Eq. (112) of Ref. \cite{Sasaki_2003}.

  \textcolor{colour1}{ \uppercase\expandafter{\romannumeral2}. Reissner-Nordstr{\"{o}}m BHs}

The potential $V(r)$ of Reissner-Nordstr{\"{o}}m BHs for the massive charged scalar perturbation field \cite{Garc_a_2021} is
 \begin{equation}
   {V_{{\rm{RN}}}} = {{(\omega r^2-e Q r)}^2} -\Delta_2 \left( {{\lambda} + {\mu ^2}{r^2}} \right),
 \end{equation}
 with $r_{\pm}=M \pm \sqrt {{M^2} - {Q^2}}$. Here, $Q$ is the charge of the black hole and $e$ is the elementary charge.

 \textcolor{colour1}{ \uppercase\expandafter{\romannumeral3}. high-dimensional Schwarzschild BHs}

The potential $V(r)$ of the background of a ($4 + 1$)-dimensional, non-rotating, neutral BHs projected onto a 3-brane with the perturbation fields of $s=0, \frac{1}{2}, 1$ \cite{Harris_2003}, that is
\begin{equation}
\begin{array}{*{20}{l}}
{{V_{(4 + 1)}} = {\omega ^2}{r^4} - is\omega \hat r_{\rm{H}}^2r}\\
{\quad \quad  + {\Delta _2}\left( {2i\omega sr + s{\Delta ^{\prime \prime }} - 2s - \lambda } \right)},
\end{array}
\end{equation}
with ${r_\pm} =\pm  {\hat r}_{\rm H}$. The horizon radius $ {\hat r}_{\rm H}$ is shown in Eq. (2.4) of Ref. \cite{Harris_2003}.

 \textcolor{colour1}{ \uppercase\expandafter{\romannumeral4}. Kerr BHs}

The potential  $V(r)$ of  Kerr BHs for all perturbation fields \cite{Teukolsky1973} is
\begin{equation}
  V_{\rm K} = {K^2} - isK\Delta_2  ' + \Delta_2  \left( {2isK' - \lambda } \right),
\end{equation}
where ${r_\pm} = M \pm \sqrt {{M^2} - {a^2}} $ and $K = \left( {{r^2} + {a^2}} \right)\omega  - ma$.

\textcolor{colour1}{ \uppercase\expandafter{\romannumeral5}. Sixth-derivative correction to Kerr BHs}

 $V(r)$ of sixth-derivative correction to Kerr BHs for all perturbation fields \cite{Cano_2019,Cano_2023} is
\begin{align}
 {V_{{\rm{K6}}}}& = {K^2} + \Delta_2 \left( {isK' - \hat \lambda } \right)  \nonumber \\
 &- is\left(2M\omega ( {{r^2} - {a^2}} ) - am{{\Delta'_2}}\right)
\end{align}
where $\hat \lambda  = {a^2}{\omega ^2} - s + {B_{lm}}$ and  ${B_{lm}}$ is the angular separation constant in Refs. \cite{Cano_2019,Cano_2023}.

  \textcolor{colour1}{ \uppercase\expandafter{\romannumeral6}. Kerr-Sen BHs}

The potential $V(r)$ of  Kerr-Sen BHs \cite{Sen_1992} for the massive charged scalar perturbation field \cite{Wu_2003,Siahaan_2015,Bernard_2016} is
\begin{equation}\label{eq:Vks}
\begin{array}{l} V_{\rm KS} = {(\omega (\Delta_2  + 2Mr) - eQr - am)^2}\\
\quad \quad - {\Delta _2}\left( {{\mu ^2}(\Delta_2  + 2Mr) + {\lambda } } \right),
\end{array}
\end{equation}
where ${r_\pm} = M-b \pm \sqrt {{(M-b)^2} - {a^2}} $ and $b = Q^2/2M$.
\begin{table*}[htbp]
	\centering
\caption{ General Form of Teukolsky Equations }\label{table:GFTE}
      \begin{tabular}{lccccc}
  \toprule
Potential Term $V(r)$     &   $ \{{r_\pm}\}$      & Perturbation Fields          \\   \midrule
${V_{{\rm{Sch}}}} = {\omega ^2}{r^4} + 2is\omega {r^2}(r - 3M) - {\Delta _2}{\lambda }$ & $ \left\{M \pm M\right\}$  &  all spins      \\  \midrule
$ {V_{{\rm{RN}}}} = {{(\omega r^2-e Q r)}^2} -\Delta_2 \left( {{\lambda} + {\mu ^2}{r^2}} \right)$ &  $ \left\{M \pm \sqrt {{M^2} - {Q^2}}\right\}$ &  $s=0$ \\  \midrule
 $\begin{array}{*{20}{l}}
{{V_{(4 + 1)}} = {\omega ^2}{r^4} - is\omega \hat r_{\rm{H}}^2r}\\
{\quad \quad  + {\Delta _2}\left( {2i\omega sr + s{\Delta ^{\prime \prime }} - 2s - \lambda } \right)}
\end{array}$   & $\{ {\pm {\hat r}_{\rm H}}\} $      &  $|s|=0, \frac{1}{2}, 1 \quad $    \\  \midrule
$ V_{\rm K} = {K^2} - isK\Delta_2  ' + \Delta_2  \left( {2isK' -\lambda } \right)$  &   $ \left\{ {M \pm \sqrt {{M^2} - {a^2}} } \right\}$   &  all spins \\  \midrule
$ \begin{aligned}
 {V_{{\rm{K6}}}}& = {K^2} + \Delta_2 \left( {isK' - \hat \lambda } \right)  \nonumber \\
 &- is\left(2M\omega ( {{r^2} - {a^2}} ) - am{{\Delta_2 '}}\right)
\end{aligned}$  &   $ \left\{ {M \pm \sqrt {{M^2} - {a^2}} } \right\}$   &  all spins \\\midrule
 $\begin{array}{l} V_{\rm KS} = {(\omega (\Delta_2  + 2Mr) - eQr - am)^2}\\
\quad \quad - {\Delta _2}\left( {{\mu ^2}(\Delta_2  + 2Mr) + {\lambda } } \right)
\end{array}$   &   $ \left\{ {M - b \pm \sqrt {{{(M - b)}^2} - {a^2}} } \right\}$     &  $s=0$  \\\midrule
 ${V_{{\rm{KN1}}}} = {K^2} - {\rm{i}}sK\Delta_2 ' + \Delta_2 \left( {2{\rm{i}}sK' - \lambda } \right)$    & {$ \left\{ {M \pm \sqrt {{M^2} - {a^2} - {Q^2}} } \right\}$}   &  all spins \\  \midrule
${V_{{\rm{KN2}}}} =K^2-\lambda{\Delta_{2}}$  &  $ \left\{ {M \pm \sqrt {{M^2}-{a^2}-{Q^2}} } \right\}$   &   all spins \\  \midrule
 $V_{\rm KN3}= {\left( {K - eQr} \right)^2} - \Delta_2  \left( {{\mu ^2}\left( {{r^2} + {a^2}} \right) + \lambda } \right)\quad $    & $ \left\{ {M \pm \sqrt {{M^2} - {a^2} - {Q^2}} } \right\}$ &  $s=0$ \\  \midrule
 $\begin{array}{l}
{V_{{\rm{KN4}}}} = {\left( {K - eQr} \right)^2} - is\left( {K - eQr} \right){\Delta'_{2}}\\
\quad \quad \quad + {\Delta _2}\left( {2{\rm{i}}sK' + \lambda } \right)\end{array}$ &$ \left\{ {M \pm \sqrt {{M^2} - {a^2} - {Q^2}} } \right\}$ &    $|s|=1,2$\\
\bottomrule
\end{tabular}
\end{table*}

 \textcolor{colour1}{ \uppercase\expandafter{\romannumeral7}. Kerr-Newman BHs}

For the curvature perturbation equation, due to the coupling between different types of perturbation fields, it seems impossible to transform the general perturbation of KN BHs into a single equation except for some limit cases or the scalar perturbation field. We collected four examples with ${r_ \pm } = M \pm \sqrt {{M^2} - {a^2} - {Q^2}}$ as follow:

\textcolor{colour1}{ \uppercase\expandafter{\romannumeral7}-1.} Using weakly charged approximation, Dudley and Finley \cite{DF1979JMP,Dudley_1977PRL} derived a Teukolsky-like equation (named as Dudley-Finley (DF) equation) for all spin fields. And $V(r)$ of DF equation for all perturbation fields is

\begin{equation}\label{eq:Vkn1}
  {V_{{\rm{KN1}}}} = {K^2} - {\rm{i}}sK\Delta_2 ' + \Delta_2 \left( {2{\rm{i}}sK' - \lambda } \right),
\end{equation}
While the QNMs of DF equation yield exact solutions for scalar perturbations, they are considered a conceptually questionable approximation for gravitational and electromagnetic modes\cite{Berti_2005,Mark_2015cyb}.

\textcolor{colour1}{ \uppercase\expandafter{\romannumeral7}-2.} $V(r)$ of the eikonal limit $\ell \gg1$ for all perturbation fields \cite{Li_2021} is
\begin{equation}\label{eq:Vkn2}
  {V_{{\rm{KN2}}}} =K^2-\lambda{\Delta_{2}},
\end{equation}

\textcolor{colour1}{ \uppercase\expandafter{\romannumeral7}-3.} $V(r)$ of  Kerr-Newman BH for charged massive scalar perturbation field \cite{Hod_2014,Hod_2015} is
\begin{equation}\label{eq:Vkn3}
V_{\rm KN3}= {\left( {K - eQr} \right)^2} - \Delta_2  \left( {{\mu ^2}\left( {{r^2} + {a^2}} \right) + \lambda } \right),
\end{equation}

\textcolor{colour1}{ \uppercase\expandafter{\romannumeral7}-4.} $V(r)$ of  Kerr-Newman BH for charged photons and gravitons perturbation fields \cite{Hartman_2010} is
\begin{equation}\label{eq:Vkn4}
\begin{array}{l}
{V_{{\rm{KN4}}}} = {\left( {K - eQr} \right)^2} - is\left( {K - eQr} \right){\Delta'_{2}}\\
\quad \quad \quad + {\Delta _2}\left( {2{\rm{i}}sK' + \lambda } \right),
\end{array}
\end{equation}

All the black holes given above satisfy Type-D black holes, that is, the four Weyl tensors (${\Psi _0},{\Psi _1},{\Psi _3},{\Psi _4}$) are 0. The potential terms of the above ten Teukolsky equations are summarized in \Cref{table:GFTE}.

\section{Fluxes and Green's Function Method}\label{sec:Flux-GFM}
Teukolsky equations with the source have found widespread application in the study of Radiation fluxes and GW waveforms \cite{Sasaki_2003,Poisson:2011,Pound:2021,Amaro_Seoane_2018,AmaroSeoane2020,Isoyama_2022,Fujita:2020}, absorption \cite{Teukolsky1974Perturbations} of gravitational waves and tail correction \cite{Casals_2015} of linear field perturbations, as well as scalar self-force \cite{Warburton_2010} and electromagnetic self-force \cite{Torres_2022} acting on a charged particle in Kerr spacetime.
In this section, the Green's function method is employed to solve the Teukolsky equation with the source, in order to derive radiation fluxes.
Using the Green function method, it is necessary to construct  two homogeneous solutions of \cref{eq:GFoTRE}, denoted as $R^{\rm in}_{\ell\omega}(r)$ and $R^{\rm up}_{\ell\omega}(r)$, that satisfy the following boundary conditions:
\begin{align}
    & R_{\ell m\omega }^{{\rm{in}}}\to \left\{ {
    \begin{array}{*{20}{l}}
{B_{\ell m\omega }^{{\rm{trans}}}{\Delta ^{ - s}}{{\rm{e}}^{ - iP{r^*}}},}&{r \to {r_ + }}\\
{{B_{\ell m\omega }^{{\rm{ref}}}}{r^{1-2s}}{{\rm{e}}^{i\omega {r^*}}} + {{B_{\ell m\omega }^{{\rm{inc}}}}}r^{-1}{{\rm{e}}^{ - i\omega {r^*}}},}&{r \to  + \infty ,}
\end{array}
    } \right.\label{eq:boundary1}\\
 & R_{\ell m\omega }^{{\rm{up}}} \to \left\{ {\begin{array}{*{20}{l}}
{C_{\ell m\omega }^{{\rm{up}}}{{\rm{e}}^{iP{r^*}}} + {C_{\ell m\omega }^{{\rm{ref}}}}\Delta ^{-s}{{\rm{e}}^{ - iP{r^*}}},}&{r \to {r_ + },}\\
{{C_{\ell m\omega }^{{\rm{trans}}}}{r^{1-2s}}{{\rm{e}}^{i\omega {r^*}}},}&{r \to  + \infty ,}
\end{array}} \right.\label{eq:boundary2}
\end{align}
where $P = P(\omega, a, m, r_{\pm}, Q, \mu)$ and $\Delta_2$ is abbreviated as $\Delta$. The tortoise coordinate $r^*$ can be defined by
\begin{align}\label{}
  {r^*} &= r + \frac{{{r_ + } + {r_ - }}}{{{r_ + } - {r_ - }}}{r_ + }\ln \frac{{r - {r_ + }}}{{{r_ + } + {r_ - }}} \nonumber \\
  &- \frac{{{r_ + } + {r_ - }}}{{{r_ + } - {r_ - }}}{r_ - }\ln \frac{{r - {r_ - }}}{{2M}}.
\end{align}

With the two homogeneous solutions $R_{\ell m\omega }^{{\rm{in,up}}}$, we can easily obtain the solution of the inhomogeneous Teukolsky equation  \eqref{eq:GFoTRE} with purely ingoing behavior at the horizon and purely outgoing behavior at infinity, which is described by
\begin{equation}\label{eq:Rlmw-source}
{R_{\ell m\omega }}(r) ={}_{s} \tilde Z_{\ell m\omega }^{\infty}R_{\ell m\omega }^{{\rm{up}}}(r) + {}_{s}\tilde Z_{\ell m\omega }^{{\rm{H}}}R_{\ell m\omega }^{{\rm{in}}}(r),
\end{equation}

The asymptotic amplitudes $\tilde Z_{\ell m\omega }^{{\infty,\rm{H}},s}$ are radial integrals over the source term:
\begin{equation}\label{eq:Z0}
 {}_s \tilde Z_{\ell m\omega }^{{\infty,\rm{H}}} = \frac{1}{{W_{\rm C}}}\int_{{r_ + }}^\infty  d r'\frac{{{{}_{s}T_{\ell m\omega }}(r')R_{\ell m\omega }^{{\rm{in,up}}}(r')}}{{{\Delta ^2}(r')}},
\end{equation}
where ${{\rm{W}}_C}$ is the conserved Wronskian, that is
\begin{align}\label{}
   {W_{\rm C}} &= R_{\ell m\omega }^{{\rm{up}}}\frac{d}{{d{r^*}}}R_{\ell m\omega }^{{\rm{in}}} - R_{\ell m\omega }^{{\rm{in}}}\frac{d}{{d{r^*}}}R_{\ell m\omega }^{{\rm{up}}} \nonumber \\
  &= 2i\omega B_{\ell m\omega }^{{\rm{inc}}}C_{\ell m\omega }^{{\rm{trans}}}.
\end{align}

The amplitudes ${}_s \tilde Z_{\ell m\omega}^{{\rm H},\infty}$ of inhomogeneous equations \eqref{eq:GFoTRE} can be employed to precisely obtain energy and angular momentum fluxes for scalar, electromagnetic, and gravitational  perturbations.
In the case of circular orbits, ${}_s \tilde Z_{\ell m\omega}^{{\rm H},\infty}$ in \cref{eq:Z0} takes the form
\begin{equation}
  {}_s \tilde Z_{\ell m\omega}^{{\rm H},\infty}= {}_s \tilde Z_{\ell m\omega}^{{\rm H},\infty}\,\delta(\omega-m\,\Omega).
\label{eq:tildeZ}
\end{equation}

The time-averaged luminosity (energy flux) at infinity is provided by \cite{Teukolsky1974Perturbations}
\begin{equation}\label{eq:dEdtInf}
 {\textcolor{white}{\biggl{|}}}_s{}\left\langle {\frac{{dE}}{{dt}}}\right\rangle _\infty = \sum\limits_{\ell  = 2}^\infty  {\sum\limits_{m =  - \ell }^\ell  {_s{\beta _{\ell m\omega }}} } \frac{{|{}_s \tilde Z_{\ell m\omega}^{\infty}{|^2}}}{{4\pi {\omega ^{2(|s| - 1)}}}},
\end{equation}
where $\left<\cdots\right>$ represents the time average and  $\omega =m\Omega$.

Similarly, the time-averaged luminosity at the horizon becomes\cite{Teukolsky1974Perturbations,Fujita_2015}
\begin{equation}\label{eq:dEdtH}
{\textcolor{white}{\biggl{|}}}_s{}\left\langle {\frac{{dE}}{{dt}}} \right\rangle _{\rm{H}} = \sum\limits_{\ell  = 2}^\infty  {\sum\limits_{m =  - \ell }^\ell  {{}_s{\alpha _{\ell m\omega }}} } \frac{{|{}_s Z_{\ell m\omega }^{{\rm{H}}}{|^2}}}{{4\pi {\omega ^2}}},
\end{equation}
where $\alpha^s_{\ell m\omega}$ and $\beta^s_{\ell m\omega}$ are the coefficients that contain the Starobinsky-Teukolsky constant \cite{Press1973,Fujita_2015GSF}.

 The angular momentum flux at infinity and event horizon for our case are given by
\begin{equation}\label{eq:dJdt}
 {\textcolor{white}{\biggl{|}}}_s{}{\left\langle {\frac{{dJ}}{{dt}}} \right\rangle _{{\rm{H,}}\infty }} = \frac{1}{\Omega }{{\textcolor{white}{\biggl{|}}}_s\left\langle {\frac{{dE}}{{dt}}} \right\rangle _{{\rm{H,}}\infty }}
\end{equation}

%

Moreover, the paper presents several functions with symmetry, which we summarize as follows:
\begin{subequations}
  \begin{align}
&{}_s Z_{\ell , - m\omega }^{\infty ,{\rm{H}}} = {( - 1)^\ell }{}_s\bar{Z}_{\ell m\omega }^{\infty ,{\rm{H}}},\\
& {{} _s}{S_{\ell m}}(\theta ) = {( - 1)^{(s + \ell )}}{{\kern 1pt} _s}{S_{\ell , - m}}(\pi  - \theta ),\\
& {{\bar R}_{\ell , - m, - \omega }} = {R_{\ell m\omega }},
\end{align}
\end{subequations}
where the bar denotes complex conjugation.

\section{Analytical Solution of Teukolsky Equations without the source}\label{sec:GsolGFRTE}
Teukolsky equations without the source have been extensively utilized to study various physical phenomena related to black holes, such as quasinormal modes \cite{leaver1985a,Cook_2014,Berti_2005,Mark_2015cyb,Li_2021}, Hawking radiation \cite{Harris_2003}, near-superradiant scattering \cite{Hartman_2010}, scalar clouds \cite{Siahaan_2015,Bernard_2016,Hod_2014,Hod_2015}, the central engine \cite{fiziev2009new} of Gamma-ray bursts, and cosmic jets \cite{Staicova_2010}. The homogenous radial Teukolsky equations (HRTEs) can be expressed as
\begin{equation}\label{eq:HRTEs}
 \left[ {\Delta^{ - s + 1}\frac{d}{{dr}}\Delta^{s + 1}\frac{d}{{dr}} + V(r)} \right]{R_{\ell m\omega }} = 0,
\end{equation}
\subsection{General Solution of HRTEs}

\cref{eq:HRTEs} is a second-order ordinary differential equation (ODE) whose general solution can be expressed as:
\begin{equation}
{R_{\ell m\omega }}(r) = {C_1}R_0^\beta (r) + {C_2}R_0^{ - \beta }(r), \label{eq:D2Gsol}
\end{equation}
where $R_0^{\pm\beta }(r)$ are two linear independent particular solutions of \cref{eq:HRTEs}, $C_1$ and $C_2$ are constants that should be determined based on different boundary conditions.

The HRTEs \eqref{eq:HRTEs} is not the standard ODE that matches the known special function.
By utilizing the so-called S-homotopic transformation \cite{slavyanov2000special}, the HRTEs \eqref{eq:HRTEs} can be transformed into an ODE that corresponds to a special  function known as the Heun class equation.
Thus, two linear independent particular solutions of \cref{eq:HRTEs} over the entire range $r\in {\cal{R}}= [r_H,\infty)$ can be obtained as
\begin{equation}\label{eq:D2Psol}
R_0^{ \pm \beta }(r) = S_{\rm{0}}^{\pm\beta} (x)\mathbb{H}^{ \pm \beta }(x),
\end{equation}
where $x$ is defined as a new coordinate that is obtained by applying a M\"{o}bius (isomorphic) transformation, which is
\begin{equation}\label{eq:mtran}
x =- \frac{{  r - r_+}}{{r_ + } - {r_ - }}.
\end{equation}

Introducing the unnormalized S-homotopic transformation
\begin{equation}\label{eq:D2SHT}
  S_{\rm{0}}^{\pm \beta}(x) = {\left( { - x} \right)^{ \frac{1}{2}(\pm \beta-s) }}{\left( {1 - x} \right)^{\frac{1}{2}(\gamma-s) }}{{\rm{e}}^{\frac{1}{2}\alpha x}},
\end{equation}
which can be regarded as the  asymptotic behaviors of the function $R_0^{ \pm \beta }(r)$  at regular singularities, then substituting \Cref{eq:D2Psol,eq:mtran,eq:D2SHT} into \cref{eq:HRTEs}, we can obtain standard confluent Heun equation
\begin{align}\label{eq:HeunC}
&\mathbb{H} '' - \frac{{\left( { - {x^2}\alpha  + \left( { - 2 - \beta  - \gamma  + \alpha } \right)x + 1 + \beta } \right)}}{{x\left( {x - 1} \right)}}\mathbb{H}'\nonumber\\
 &  - \left( \begin{array}{l}
\left( {\left( { - 2 - \beta  - \gamma } \right)\alpha  - 2\delta } \right)x \\
 + \left( {\beta  + 1} \right)\alpha  + \left( { - \gamma  - 1} \right)\beta  - \gamma  - 2\eta
\end{array} \right)\frac{\mathbb{H} }{{2x\left( {x - 1} \right)}} = 0.
\end{align}

Two linear independent particular solutions $\mathbb{H}_{\rm{0}}^{\pm \beta } (x)$ of \cref{eq:HeunC} can be expressed as
\begin{subequations}\label{eq:HeunC-parsol}
\begin{align}
&\mathbb{H}_{\rm{0}}^{\beta } (x) = {\rm{HeunC}}(\alpha , \beta ,\gamma ,\delta ,\eta ;x) ,\\
&\mathbb{H}_{\rm{0}}^{-\beta} (x) = {( - x)^{ - \beta }}{\rm{HeunC}}(\alpha , - \beta ,\gamma ,\delta ,\eta ;x),
\end{align}
\end{subequations}
where  $\text{HeunC}$ is the confluent Heun\footnote{In this paper, the HeunC function are implemented using the corresponding functions and symbolic notations provided by the computational software \textit{Maple}.}  function \cite{Ronveaux:1995,slavyanov2000special,olver2010nist}, and $\alpha,\beta,\gamma,\delta$ and $\eta$ are parameters that should be determined for given black holes.
In \Cref{table:parameter}, we calculate the parameters corresponding to the ten Teukolsky equations proposed in \Cref{sec:GTF} .


\begin{table*}[htbp]
  \centering
\caption{ parameters that should be determined for given black holes}\label{table:parameter}
\begin{threeparttable}
    \begin{tabular}{c|c|c|c|c|c} \toprule
 $V(r)$   &   $\alpha$&   $\beta$& $\gamma$& $\delta$ & $\eta$ \\ \midrule
${V_{{\rm{Sch}}}} $ &   $ 2i\omega {r_+}$  &  $ - s - 2i\omega {r_+}$  & $ s$  &  $ -2i s\omega {r_ + } - 2{\omega ^2}r_ + ^2$   &  $\begin{array}{l}
2is\omega {r_ + } + 2{\omega ^2}r_ + ^2\\
 - \frac{1}{2}{s^2} - s - \lambda
\end{array}$     \\ \midrule
$V_{{\rm{RN}}} $ &   $ - 2\sqrt {{\mu ^2} - {\omega ^2}} {r_{\rm{x}}}$
&  $\frac{{2i{r_ + }}}{{{r_{\rm{x}}}}}(\omega {r_ + } - eQ)$
& $ \frac{{2i{r_ - }}}{{{r_{\rm{x}}}}}(eQ - \omega {r_ - })$
&  $\begin{array}{l}2r_{\rm{x}}^2{\omega ^2} - 2eQ\omega {r_{\rm{x}}}\\- {\mu ^2}(r_ - ^2 - r_ + ^2){r_{\rm{x}}}\end{array}$
&   $\begin{array}{l}
 - \lambda  - \frac{{{r_ + }}}{{r_{\rm{x}}^2}}(2{Q^2}{e^2}{r_ - } + \\
2Qe\omega r_ + ^2 - 2{\omega ^2}r_ + ^3\\
 + {\mu ^2}r_ - ^2{r_ + } - 2{\mu ^2}{r_ - }r_ + ^2\\
 + {\mu ^2}r_ + ^3 + 4{\omega ^2}{r_ - }r_ + ^2\\
 - 6Qe\omega {r_ - }{r_ + })
\end{array}$     \\ \midrule
$V_{(4 + 1)}$ &   $- 2i\omega {r_{\rm{x}}}$
&  $\begin{array}{l}
 - \frac{1}{{{r_{\rm{x}}}}}(r_{\rm{x}}^2{s^2} + \\
4is\omega r_ + ^3 - 4{\omega ^2}r_ + ^4{)^{\frac{1}{2}}}
\end{array}$
& $\begin{array}{l}
\frac{1}{{{r_{\rm{x}}}}}(r_{\rm{x}}^2{s^2} + \\
4is\omega {r_ - }r_ + ^2 - 4{\omega ^2}r_ - ^4{)^{\frac{1}{2}}}
\end{array} $
&  $2\omega {r_{\rm{x}}}\left( {\omega {r_{\rm{x}}} + is} \right)$
&   $ \begin{array}{l}
 - \frac{{{s^2}}}{2} + 3s - \lambda  + \\
\frac{{\omega {r_ + }}}{{r_{\rm{x}}^2}}(2isr_ - ^2 - 3is{r_ - }{r_ + } + \\
3isr_ + ^2 - 4\omega {r_ - }r_ + ^2 + 2\omega r_ + ^3)
\end{array}$     \\ \midrule

$V_{\rm K} $ &   $- 2i\omega r_{\rm x}$
&  $-s+\frac{{2i\omega (r_ + ^2 + {a^2}) - 2iam}}{{{r_{\rm{x}}}}}$
& $ s+\frac{{2i\omega (r_ - ^2 + {a^2}) - 2iam}}{{{r_{\rm{x}}}}}$
&  $2\omega  r_{\rm x}( {\omega r_{\rm{w}}+ is})$
&   $\begin{array}{l} 2is\omega {r_ + }- \frac{1}{2}{s^2} - s - \lambda \\ - \frac{2}{{r_{\rm{x}}^2}}( {\omega r_{\rm m}  - am} ) ({\omega r_{\rm o}  - am})  \end{array}$     \\ \midrule


$V_{\rm K6} $ &$- 2i\omega r_{\rm x}$
&$\begin{array}{l}
 - \frac{1}{{{r_{\rm{x}}}}}\{ r_{\rm{x}}^2{s^2} - 4r_{\rm{o}}^2{\omega ^2}\\
 + 8is[am(M - {r_ + })\\
 - M\omega ({a^2} - r_ + ^2)]\\
 + 16Mam\omega {r_ + }\\
 - 4{m^2}{a^2}{\} ^{\frac{1}{2}}}
\end{array}$
&$\begin{array}{l}
 - \frac{1}{{{r_{\rm{x}}}}}\{ r_{\rm{x}}^2{s^2}\\
 - 4{\left( {{a^2} + r_ - ^2} \right)^2}{\omega ^2}\\
 + 8is[am(M - {r_ - })\\
 - M\omega ({a^2} - r_ - ^2)]\\
 + 16Mam\omega {r_ - }\\
 - 4{m^2}{a^2}{\} ^{\frac{1}{2}}}
\end{array}$
& $2\omega {r_{\rm{x}}}\left( {\omega r_{\rm{w}} + is} \right)$
&   $\begin{array}{l}
2is\omega {r_ + } - s - \frac{1}{2}{s^2} - \hat \lambda \\
 + \frac{1}{{r_{\rm{x}}^2}}\{  - 2{a^4}{\omega ^2}- 2(2iMs\omega \\
 + 2{r_ - }{r_ + }{\omega ^2} + {m^2}){a^2}\\
 + ma[4\omega Mr_{\rm{w}}+ 2is(2M - r_{\rm{w}})]\\
 + 2\omega {r_ + }[2iMs{r_ - }\\
 - \omega r_ + ^2(2{r_ - } - {r_ + })]\}
\end{array}$     \\\midrule

$V_{\rm KS} $ &   $ - 2\sqrt {{\mu ^2} - {\omega ^2}} {r_{\rm{x}}}$
&  $  \begin{array}{l}
 - \frac{{2i}}{{{r_{\rm{x}}}}}(2M\omega{r_ + } \\
 - ma - eQ{r_ + })
\end{array}$
& $\begin{array}{l}
\frac{{2i}}{{{r_{\rm{x}}}}}(2M\omega {r_ - }\\
 - ma - eQ{r_ - })
\end{array}$
&  $ \begin{array}{l}
 - 2{r_{\rm{x}}}(eQ\omega \\
 + ({\mu ^2} - 2{\omega ^2})M)
\end{array}$
&   $ \begin{array}{l}
 - 2{r_ + }({\mu ^2}M + 2{\omega ^2}M + eQ\omega )\\
 - \lambda  - 2am\omega  + \frac{1}{{r_{\rm{x}}^2}}\left\{ {[ - 2{a^2}{m^2}} \right.\\
 + 2{r_ - }{r_ + }(4M\omega  - eQ)\\
 - 2ma({r_ - } + {r_ + })]eQ\\
 - 8M\omega ( - \omega r_ + ^3 + 2\omega {r_ - } r_ + ^2\\
 + M\omega {r_ - }{r_ + } - {r_ + } r_ - ^2\omega \\
\left. { - \frac{{ma}}{2}r_{\rm{w}})} \right\}
\end{array}$     \\ \midrule

$V_{\rm{KN1}} $ &   $- 2i\omega r_{\rm x}$
&  $-s+\frac{{2i\omega (r_ + ^2 + {a^2}) - 2iam}}{{{r_{\rm{x}}}}}$
& $ s+\frac{{2i\omega (r_ - ^2 + {a^2}) - 2iam}}{{{r_{\rm{x}}}}}$
&  $2\omega  r_{\rm x}( {\omega ( r_{-}+r_{+})+ is})$
&   $\begin{array}{l} 2is\omega {r_ + }- \frac{1}{2}{s^2} - s - \lambda \\ - \frac{2}{{r_{\rm{x}}^2}}\left[ {\omega r_{\rm m}  - am} \right] \times \\
\left[ {\omega r_{\rm o}  - am} \right] \end{array}$     \\ \midrule

$V_{\rm{KN2}} $ &   $- 2i\omega r_{\rm x}$
&  $\begin{array}{l}
 - \frac{2}{{{r_{\rm{x}}}}}[r_{\rm{x}}^2{s^2} - \\
{( - ma + {r_{\rm{m}}}\omega )^2}{]^{\frac{1}{2}}}
\end{array}$
& $ \begin{array}{l}
\frac{2}{{{r_{\rm{x}}}}}[r_{\rm{x}}^2{s^2} - \\
{( - ma + {r_{\rm{m}}}\omega )^2}{]^{\frac{1}{2}}}
\end{array}$
&  $2{\omega ^2}r_ - ^2 - 2{\omega ^2}r_ + ^2$
&   $\begin{array}{l}- \frac{1}{2}{s^2} - s - \lambda \\ - \frac{2}{{r_{\rm{x}}^2}}\left[ {\omega r_{\rm m}  - am} \right] \times \\
\left[ {\omega r_{\rm o}  - am} \right] \end{array}$     \\ \midrule

$V_{\rm{KN3}} $ &   $ - 2\sqrt {{\mu ^2} - {\omega ^2}} {r_{\rm{x}}}$
&  $ \begin{array}{l}- \frac{{2{\kern 1pt} i}}{{{r_{\rm{x}}}}}({a^2}\omega  + \omega r_ + ^2\\- ma - eQ{r_ + })\end{array}$
& $\begin{array}{l}\frac{{2{\kern 1pt} i}}{{{r_{\rm{x}}}}}({a^2}\omega  + \omega r_ - ^2\\- ma - eQ{r_ - })\end{array}$
&  $ \begin{array}{l} - {r_{\rm{x}}}[2eQ\omega  + \\({r_ - } + {r_ + })({\mu ^2} - 2{\omega ^2})]\end{array}$
&   $ \begin{array}{l}
 - {\mu ^2}r_{\rm m}- \lambda \\
 - \frac{2}{{r_{\rm{x}}^2}}(\omega r_{\rm m}  - ma - {r_ +}eQ) \\
\times (  \omega r_{\rm o} - ma - eQ{r_-})
\end{array}$     \\ \midrule

$V_{\rm{KN4}} $ &   $- 2i\omega r_{\rm x}$
&  $\begin{array}{l}
 - \frac{1}{{r_{\rm{x}}^2}}[ - 4{a^4}{\omega ^2}\\
 + 8{a^3}m\omega \\
 + 4{a^2}(2eQ\omega {r_ + }\\
 - is\omega {r_{\rm{x}}}\\
 - 2{\omega ^2}r_ + ^2 - {m^2})\\
 + 4ma(2i\omega r_ + ^2\\
 - 2eQ{r_ + } + s{r_{\rm{x}}})\\
 + r_{\rm{x}}^2{s^2} - 4is\omega r_ + ^2r_{\rm{x}}^2\\
 - 4r_ + ^2{\left( {eQ - \omega {r_ + }} \right)^2}{]^{\frac{1}{2}}}
\end{array}$
& $\begin{array}{l}
\frac{1}{{{r_{\rm{x}}}}}[ - 4{a^4}{\omega ^2}\\
 + 8{a^3}m\omega \\
 + 4{a^2}(2eQ\omega {r_ - }\\
 + is\omega {r_{\rm{x}}}\\
 - 2{\omega ^2}r_ - ^2 - {m^2})\\
 - 4ma(is{r_{\rm{x}}}\\
 + 2eQ{r_ - } - 2\omega r_ - ^2)\\
 + r_{\rm{x}}^2{s^2} + 4is\omega r_ - ^2{r_{\rm{x}}}\\
 - 4r_ - ^2{\left( {eQ - \omega {r_ - }} \right)^2}{]^{\frac{1}{2}}}
\end{array}$
&  $\begin{array}{l}
2\omega {r_{\rm{x}}}[({r_ - } + {r_ + })\omega \\
 + is - eQ]
\end{array}$
&   $\begin{array}{*{20}{l}}
{2is\omega {r_ + } - \frac{1}{2}{s^2} - s - \lambda }\\
{ - \frac{2}{{r_{\rm{x}}^2}}(\omega {r_{\rm{m}}} - ma - {r_ + }eQ)}\\
{ \times (\omega {r_{\rm{o}}} - ma - eQ{r_ - })}
\end{array}$     \\

\bottomrule

    \end{tabular}%
\begin{tablenotes}
        \footnotesize
        \item[$\clubsuit$] In \Cref{table:parameter}, some symbols are defined as: ${r_{\rm{x}}} = {r_ - } - {r_ + }$, ${r_{\rm{w}}} = {r_ - } + {r_ + }$, $r_{\rm o}={a^2} + 2{r_-} {r_+} - r_{+}^2$, $r_{\rm m} = r_+ ^2+{a^2}$.
\end{tablenotes}
\end{threeparttable}

\end{table*}

Using the $\rm HeunC$ function (\ref{eq:HeunC-parsol}) and the S-homotopic transformation (\ref{eq:D2SHT}), the general solution of \cref{eq:HRTEs} for $\Delta_2$-type can be expressed as:
\begin{align}\label{eq:D2GSol1}
{R_{\ell m\omega }} &= {C_1}S_0^\beta (x){\rm{HeunC}}(\alpha ,\beta ,\gamma ,\delta ,\eta ;x) \nonumber\\
&+ {C_2}S_0^{ - \beta }(x){\rm{HeunC}}(\alpha , - \beta ,\gamma ,\delta ,\eta ;x).
\end{align}

The general solution \eqref{eq:D2GSol1} can be effectively applied to a specific physical model by fixing the parameters ($\alpha,\beta,\gamma,\delta$, $\eta$) for a given black hole. This allows us to tailor the solution to accurately describe the characteristics and properties of the particular black hole under consideration.
$C_1$ and $C_2$  are the combination coefficients that are explicitly solved by three significant boundary conditions (Dirichlet, Neumann, and Robbin) in this paper. The application of these boundary conditions requires the asymptotic behavior of the confluent Heun function at the outer horizon and infinity, respectively.
Expanding the confluent Heun function in power series for the independent variable $x$ around the regular singular point $x=0$ \cite{Ronveaux:1995}, yields the following asymptotic behavior at the outer horizon:
\begin{equation}\label{eq:hc-hor}
  \mathop {\lim }\limits_{x \to 0} {\rm{ HeunC}}(\alpha ,\beta ,\gamma ,\delta ,\eta ;x) = 1,\quad r\rightarrow r_+,
\end{equation}

Expanding the confluent Heun function in a sector around the irregular singular point at infinity \cite{Ronveaux:1995}, the asymptotic behavior at infinity can be expressed as:
\begin{align}\label{eq:hc-inf}
&\mathop {\lim }\limits_{\left| x \right| \to \infty } {\rm{HeunC}}(\alpha ,\beta ,\gamma ,\delta ,\eta ;x) \to \nonumber \\
&D_ \odot^\beta \;{x^{ - \frac{{\beta  + \gamma  + 2}}{2} - \frac{\delta }{\alpha }}} + D_ \otimes ^\beta {{\rm{e}}^{ - \alpha x}}{x^{ - \frac{{\beta  + \gamma  + 2}}{2} + \frac{\delta }{\alpha }}},\ \ \ \ r \to \infty,
\end{align}
where ${{  D}_{\otimes}^\beta }$ and ${{  D}_{\odot }^\beta }$ are undetermined constants. Only when the constants ${{  D}_{\otimes}^\beta }$ and ${{  D}_{\odot }^\beta }$ are known, the value of the combination coefficient $C_1$ and $C_2$  can be determined according to the boundary conditions.
Because the calculation process of the constants ${{D}_{\otimes}^\beta }$ and ${{D}_{\odot }^\beta }$ is very complicated and lengthy,
we will introduce their calculation process in detail in \Cref{sec:AsympForHeunC}.

While Fiziev proposed a general solutions similar to \cref{eq:D2GSol1} for solving QNMs \cite{Fiziev_2011a}, he was not acquainted with the constants ${{  D}_{\otimes}^\beta }$ and ${{  D}_{\odot }^\beta }$.
Therefore,  he believes that providing explicit analytical expressions for $C_1$ and $C_2$ is an unsolved difficult problem in mathematics\cite{Fiziev_2006,fiziev2009newa}.
In the past decade, Bezerra and Vieira have dedicated their efforts to the exploration of Hawking radiation in scalar fields\cite{VIEIRA201414,Bezerra_2014,VIEIRA2015576,Vieira_2020,VIEIRA2020168197}. They also constructed a general solution similar to \cref{eq:D2GSol1} for the Klein-Gordon equation. Since the decay rate $\Gamma_+$ does not involve the calculation of ${{  D}_{\otimes}^\beta }$ and ${{  D}_{\odot }^\beta }$, they can simply use asymptotic behavior \eqref{eq:hc-hor} to obtain the Hawking radiation spectrum.

Solving ${{  D}_{\otimes}^\beta }$ and ${{  D}_{\odot }^\beta }$ analytically is one of the key advantages of our general solutions, which is different from the general solutions of these literatures \cite{Fiziev_2006,fiziev2009classes,Fiziev_2011a,VIEIRA201414,Bezerra_2014,VIEIRA2015576,Vieira_2020,VIEIRA2020168197}.
It is also beneficial to extend the general solutions of homogeneous equations to inhomogeneous equations \eqref{eq:GFoTRE}.
In contrast, the general solutions presented by Fiziev and Vieira are specific cases within our broader framework. \cref{eq:D2GSol1} covers the general solution of QNM, Hawking radiation and other physical problems.
Furthermore, their general solution is unable to construct the Green's function required by the inhomogeneous equation and the outgoing wave solution $R_{\ell m\omega }^{{\rm{up}}}$. Consequently, we apply the general solution to construct  the solution of inhomogeneous equations and $R_{\ell m\omega }^{{\rm{up}}}$, thereby broadening its application scope.

\subsection{Ingoing and Outgoing Wave Solutions}\label{sec:in-outgongsol}
To get the general solution of the Teukolsky equation with the source under given boundary conditions, we should first find the ingoing wave and outgoing wave solutions $R_{\ell m\omega}^{\rm in,up}$ of the homogenous Teukolsky equation, then obtain the general solution of the inhomogeneous Teukolsky equation utilizing the Green's function method \cite{Detweiler:1978ge}.
The most common methods for solving the homogenous solutions $R^{\rm in,up}_{\ell m\omega}$  are the PN expansion \cite{Tagoshi_1994} of the SN equation and the MST method\cite{Sasaki_2003,Mano1996RWE,Mano_1996}. The former method involves utilizing the Chandrasekhar-Sasaki-Nakamura transformation  \cite{Chandrasekhar1975,Sasaki_1982a,Sasaki_1982b,Tagoshi_1994} to convert the HRTEs \eqref{eq:HRTEs} into the SN equation, and then derives the PN expansion of $R^{\rm in}_{\ell m\omega}$.
On the other hand, in the MST method, a series solution of the hypergeometric function that converges within a finite region is first constructed, and then a series solution of the Coulomb wave function that converges at infinity is generated. The two solutions are subsequently matched to yield a convergent solution that extends from the horizon to infinity.

Now, we commence with the construction of the ingoing and outgoing wave solutions from the solution \eqref{eq:D2GSol1}.
Using the asymptotic properties \eqref{eq:hc-hor} and \eqref{eq:hc-inf} of the confluent Heun function and the boundary condition of the ingoing wave, we can construct the ingoing wave solution. Noting the ingoing wave solution at the horizon is purely ingoing wave, we have $C_2^{\rm in}=0$. Thus, the ingoing wave solution $R_{\ell m\omega }^{{\rm{in}}}$ is given by
\begin{align}\label{eq:uHor}
R_{\ell m\omega }^{{\rm{in}}} &= C_1^{{\rm{in}}}R_{\rm{0}}^\beta  + C_2^{{\rm{in}}}R_{\rm{0}}^{ - \beta } = C_1^{{\rm{in}}}R_{\rm{0}}^\beta \nonumber \\
 &=C_1^{{\rm{in}}} S_0^\beta (x) {\rm{HeunC}}(\alpha ,\beta ,\gamma ,\delta ,\eta ;x).
\end{align}

Similarly, we can construct the outgoing wave solution as
\begin{align}\label{eq:uOut}
R_{\ell m\omega }^{{\rm{up}}}&= C_1^{{\rm{up}}}R_{\rm{0}}^\beta  + C_2^{{\rm{up}}}R_{\rm{0}}^{ - \beta } \nonumber \\
 &= C_1^{{\rm{up}}} S_0^\beta (x)  {\rm{HeunC}}(\alpha ,\beta ,\gamma ,\delta ,\eta ;x) \nonumber  \\
 &+ C_2^{{\rm{up}}} S_0^{-\beta} (x) {\rm{HeunC}}(\alpha ,-\beta ,\gamma ,\delta ,\eta ;x).
\end{align}

According to the asymptotic properties \eqref{eq:hc-hor} and \eqref{eq:hc-inf} of the confluent Heun function, and noting that the outgoing wave solution at infinity is purely outgoing wave, we have
\begin{align}
& C_1^{{\rm{up}}} = {\left( { - 1} \right)^{\beta+1} }\Big( {  \frac{{D_ \odot ^{ - \beta }}}{{D_ \odot ^\beta }}} \Big)C_2^{{\rm{up}}}.
\end{align}

Thus, the outgoing wave solution is described by
\begin{align}\label{eq:uOut1}
R_{\ell m\omega }^{{\rm{up}}}&= C_2^{{\rm{up}}}\Big[ {\left( { - 1} \right)^{\beta+1} }\Big( {  \frac{{D_ \odot ^{ - \beta }}}{{D_ \odot ^\beta }}} \Big)S_0^\beta (x)  {\rm{HeunC}}(\alpha ,\beta ,\gamma ,\delta ,\eta ;x) \nonumber  \\
 &+ S_0^{-\beta} (x) {\rm{HeunC}}(\alpha ,-\beta ,\gamma ,\delta ,\eta ;x)\Big].
\end{align}

Compared with PN expansion results, our ingoing wave solution \eqref{eq:uHor} is considered a complete solution without the need for series expansion. And our outgoing wave solution \eqref{eq:uOut1} is undisputedly accurate, satisfying the conservation of the Wronskian determinant for any BHs.
Additionally, the solutions \eqref{eq:uHor} and  \eqref{eq:uOut1} are not constrained by limitations stemming from slow motion and weak-field approximations, rendering its findings superior to those of the PN expansion results near the horizon.
Compared with the MST method, our method employs a special function to construct the solutions $R_{\ell m\omega }^{{\rm{in,up}}}$ which  does not involve computing two-sided infinite series or renormalized angular momentum solutions $\nu$ ( MST method faces solving the transcendental equation of renormalized angular momentum $\nu$ ). Therefore, the solutions impose no limitations, while the MST method is constrained by low-frequency approximations.

\section{Asymptotic Formula of HeunC Function at Infinity}\label{sec:AsympForHeunC}
Up to now, an analytic asymptotic expression for the confluent Heun function $\mathbb{H}(x)$ at infinity  $|x|$ has not been reported in the literature. While most literature provides a linear combination of the two asymptotic solutions of the confluent Heun function at infinity, as in \cref{eq:hc-inf}, the coefficients ${{D}_{\otimes}^\beta }$ and ${{D}_{\odot}^\beta}$ remain undetermined.
The solution $Y(x)$ to the generalized spherical wave equation (GSWE\footnote{Some researchers\cite{Figueiredo2002,El_Jaick_2013} have named GSWE as confluent Heun equations, which is not used in this paper.} ) is related to $\mathbb{H}(x)$ by $Y(x)={{\rm{e}}^{i\omega x}}\mathbb{H}(x)$.
Additionally,  $Y(x)$ can be expressed as a series in terms of Coulomb wave functions ${F_{n + \nu }}(x)$ \cite{Leaver1986} that converges for $x>0$.
Therefore, an analytic asymptotic expression for $\mathbb{H}(x)$ at infinity can also be constructed in terms of a series solution in terms of ${F_{n + \nu }}(x)$.
However, the series in terms of ${F_{n + \nu }}(x)$ does not converge at $x=0$, which prohibits normalization with the asymptotic expression of $\mathbb{H}(x)$ at $x=0$.
Instead, a proportionality relation between the two can be established, but the proportionality coefficient $\Xi$ is undetermined.
To determine this coefficient $\Xi$, we can represent $\mathbb{H}(x)$ as a series solution in terms of hypergeometric functions ${}_2{F_1}(x)$ \cite{olver2010nist}, which converge at $x=0$.
Therefore, $\mathbb{H}(x)$ can be expressed as a series in terms of ${}_2{F_1}(x)$ and as a proportionality coefficient multiplied by a series solution in terms of ${F_{n + \nu }}(x)$.
By expanding both series in the interval $0< x < \infty $, the proportionality coefficient $\Xi$ can be determined.
This mathematical technique is similar to the approach used in Refs. \cite{Mano1996RWE,Mano_1996} to determine the analytic asymptotic amplitudes of $ R_{\ell m\omega }^{{\rm{in}}}$ at infinity.

\subsection{Expansion in Series of Hypergeometric Function}
The series expansion of the hypergeometric function can be utilized to represent the confluent Heun function\footnote{This paper assumes ${\rm{Im}}(\alpha ) > 0$. }.

\begin{align}\label{eq:HeunC-2F1}
 \mathbb{H} (x)&=  {\rm{HeunC}}(\alpha ,\beta ,\gamma ,\delta ,\eta ;x)  \nonumber \\
              &= \mathbb{F} \sum\limits_{n =  - \infty }^{+\infty}  {f{{_n^{\nu} }}\,{}_2{F_1}\left( {a, { b}, { c};x} \right)}
\end{align}
with $\mathbb{F}  ={\left(\sum\nolimits_{n =  - \infty }^\infty  {f_n^\nu } \right)^{ - 1}}$ is the normalized function at $x=0$. Here, $a = n + \nu  + 1 + \frac{{\beta  + \gamma }}{2}, { b} =  - n - \nu  + \frac{{\beta  + \gamma }}{2}$ and ${ c} = \beta  + 1$.
Substituting the series solution, as given by \cref{eq:HeunC-2F1}, into the confluent Heun equation \eqref{eq:HeunC}, we can derive the following three-term recurrence relation for the expansion coefficients $f_n^\nu$.
\begin{equation}\label{recurrence-fnv}
 {{\hat \alpha }_n}f_{n + 1}^\nu  + {{\hat \beta }_n}f_n^\nu  + {{\hat \gamma }_n}f_{n - 1}^\nu  = 0
\end{equation}
where
\begin{subequations}
\begin{align}
{{\hat \alpha }_n} &=   \frac{{\left( {2n + 2\nu  + 2 - \beta  + \gamma } \right)}}{{8\left( {n + \nu  + 1} \right)\left( {2n + 2\nu  + 3} \right)}}\nonumber\\
  &\times \left( {\alpha n + \alpha \nu  + \alpha  -\delta } \right)\left( {2n + 2\nu  + 2 - \gamma  - \beta } \right),\\
{{\hat \beta }_n} &= \eta  + \frac{\delta }{2} - \frac{{{\beta ^2}}}{4} - \frac{{{\gamma ^2}}}{4} + \left( {n + \nu } \right)\left( {n + \nu  + 1} \right) \nonumber\\
 &+ \frac{{\delta \left( {\gamma  + \beta } \right)\left( {\beta  - \gamma } \right)}}{{8\left( {n + \nu } \right)\left( {n + \nu  + 1} \right)}},\\
{{\hat \gamma }_n} &= -\frac{{\left( {2n + 2\nu  + \beta  - \gamma } \right)}}{{8\left( {n + \nu } \right)\left( {2n + 2\nu  - 1} \right)}} \nonumber\\
 &\times \left( {\alpha n + \alpha \nu  + \delta } \right)\left( {2n + 2\nu  + \gamma  + \beta } \right).
\end{align}
\end{subequations}
The phase parameter $\nu$, also known as the renormalized angular momentum, may be obtained by solving a characteristic equation expressed as the sum of two infinite continued fractions \cite{Leaver1986,Figueiredo1993,Figueiredo2002,El_Jaick_2013}.
\begin{equation}\label{eq:eig}
  {{\hat \beta }_0} = \frac{{{{\hat \alpha }_{ - 1}}{{\hat \gamma }_0}}}{{{{\hat \beta }_{ - 1}} - }}\frac{{{{\hat \alpha }_{ - 2}}{{\hat \gamma }_{ - 1}}}}{{{{\hat \beta }_{ - 2}} - }}\frac{{{{\hat \alpha }_{ - 3}}{{\hat \gamma }_{ - 2}}}}{{{{\hat \beta }_{ - 3}} - }} +  \cdots
 + \frac{{{{\hat \alpha }_0}{{\hat \gamma }_1}}}{{{{\hat \beta }_1} - }}\frac{{{{\hat \alpha }_1}{{\hat \gamma }_2}}}{{{{\hat \beta }_2} - }}\frac{{{{\hat \alpha }_2}{{\hat \gamma }_3}}}{{{{\hat \beta }_3} - }} \cdots .
\end{equation}

Solving the solution $\nu$ of \cref{eq:eig} is a complex task, because \cref{eq:eig} is a transcendental equation. Nevertheless, two approaches have been developed to determine $\nu$. The first method involves presenting a series expansion of $\nu$, without having to solve the transcendental equation directly \cite{Mano1996RWE,Mano_1996,Casals_2015}. However, this approach requires enforcing low-frequency constraints. The second method, originally introduced by Fujita and Tagoshi\cite{Fujita_2004,Fujita_2005}, utilizes the Steed algorithm for continued fractions to numerically solve the transcendental equation \eqref{eq:eig} and obtain $\nu$, without requiring any constraints. This paper selects the unconstrained second method over the first due to the limitations of the series expansion approach.

The series representation of the hypergeometric function is given by \cite{olver2010nist}
\begin{equation}\label{eq:2F1series}
{}_2{F_1}({{a}},{{b}},{{c}},\tilde x) = \sum\limits_{j = 0}^\infty  {\frac{{{{( a)}_j}{{(b)}_j}}}{{{{(c)}_j}}}{{\frac{{\tilde x}}{{j!}}}^j}} ,\quad 0 < \left| {\tilde x} \right| < 1
\end{equation}
where $({{a}})_n$ denotes the Pochhammer symbol defined as $({{a}})_j={{a}}({{a}}+1)({{a}}+2)\cdots({{a}}+j-1)$ with $({{a}})_0=1$. And $\tilde x$ is new variable, defined as $\tilde x = 1/(1-x)$.

Applying the linear transformation of hypergeometric functions ( Eq. (15.3.8) in Ref. \cite{abramowitz1948handbook} ), we can derive a relation between ${}_2{F_1}(x)$ and ${}_2{F_1}( \tilde x) $.
\begin{align}\label{eq:2F1Transformation}
{}_2{F_1}&\left( {{{a}},{{b}},{{c}};x} \right) =\nonumber  \\
 &{{\tilde x}^{{a}}}\frac{{\Gamma ({{c}})\Gamma ({{b}} - {{a}})}}{{\Gamma ({{b}})\Gamma ({{c}} - {{a}})}}{{\kern 1pt} _2}{F_1}\left( {{{a}},{{c}} - {{b}},{{a}} - {{b}} + 1;\tilde x} \right) \nonumber  \\
  +& {{\tilde x}^{{b}}}\frac{{\Gamma ({{c}})\Gamma ({{a}} - {{b}})}}{{\Gamma ({{a}})\Gamma ({{c}} - {{b}})}}{{\kern 1pt} _2}{F_1}\left( {{{b}},{{c}} - {{a}},{{b}} - {{a}} + 1;\tilde x} \right).
\end{align}

The confluent Heun function can be expressed in an alternate form by utilizing \Cref{eq:HeunC-2F1,eq:2F1Transformation}, as shown below:
\begin{equation}\label{eq:HeunC-hyper}
\mathbb{H}(x)= {\mathbb{H} _{n,\nu }}\left( {\tilde x} \right) + {\mathbb{H} _{ - n, - \nu  - 1}}\left( {\tilde x} \right),
\end{equation}
where
\begin{align}
& \begin{array}{l}\label{eq:Hnv}
{\mathbb{H} _{n,\nu }}\left( {\tilde x} \right) = \mathbb{F} {{\tilde x}^{ - \nu  + \frac{{\beta  + \gamma }}{2}}}\sum\limits_{n =  - \infty }^\infty  {\frac{{\Gamma ({ {c}})\Gamma ({ {b}} - { {a}})}}{{\Gamma ({ {b}})\Gamma ({ {c}} - { {a}})}}} \\
\quad \quad \times f_n^\nu {{\kern 1pt} _2}{F_1}\left( {{ {a}},{ {c}} - { {b}};{ {a}} - { {b}} + 1;\tilde x} \right),
\end{array} \\
 & \begin{array}{l}
{\mathbb{H} _{ - n, - \nu  - 1}}\left( {\tilde x} \right) = \mathbb{F} {{\tilde x}^{\nu  + 1 + \frac{{\beta  + \gamma }}{2}}}\sum\limits_{n =  - \infty }^\infty  {\frac{{\Gamma ({ {c}})\Gamma ({ {a}} - { {b}})}}{{\Gamma ({ {a}})\Gamma ({ {c}} - { {b}})}}} \\
\quad \quad  \times f_n^{ - \nu  - 1}{{\kern 1pt} _2}{F_1}\left( {{ {b}},{ {c}} - { {a}};{ {b}} - { {a}} + 1;\tilde x} \right).
\end{array}
\end{align}

Accordingly, the recurrence relation \eqref{recurrence-fnv} possesses a structure such that ${f_n^{-\nu-1}}$ satisfies an identical recurrence relation to that of ${f_n^\nu}$.
Introducing a new coordinate variable $\hat z = 2i/(\alpha \tilde x)$ and applying \cref{eq:2F1series} into \cref{eq:Hnv} , we can expand ${\mathbb{H}_{n,\nu}}(\tilde{x})$ as a series in terms of $z$.
\begin{equation}\label{eq:hyper-expand-2}
  {\mathbb{H} _{n,\nu }}= \mathbb{F} {\left({\frac{{2i\hat z}}{\alpha }} \right)^{\nu  - \frac{{ \beta+ \gamma  }}{2}}}\sum\limits_{k =  - \infty }^\infty  {\sum\limits_{n = k}^\infty  {{C_{n,n - k}}{{\hat z}^k}} },
\end{equation}
where
\begin{align}
{C_{n,n - k}} &= \frac{{{{\left( { - n - \nu  + \frac{{\beta  + \gamma }}{2}} \right)}_{n - k}}{{\left( { - n - \nu  + \frac{\delta }{\alpha }} \right)}_{n - k}}}}{{\Gamma \left( {n + \nu  + 1 + \frac{{\beta  + \gamma }}{2}} \right)\Gamma \left( {n + \nu  + 1 + \frac{{\beta  - \gamma }}{2}} \right)}} \nonumber \\
& \times\frac{{\Gamma \left( {\beta  + 1} \right)\Gamma \left( {2n + 2\nu  + 1} \right)}}{{{{\left( { - 2n - 2\nu } \right)}_{n - k}}\left( {n - k} \right)!}}{\left( { - \frac{{i\alpha }}{2}} \right)^{ - k}}f_n^\nu,
\end{align}

\subsection{Expansion in Series of Coulomb Wave Function}\label{sec:Expan-S-CWF}

Assuming that ${\cal F}_{n,\nu }^C$ is a non-trivial solution of the confluent Heun equation \eqref{eq:HeunC}, which can be written as a series expansion of Coulomb wave functions. Here, $\hat z =   \frac{i}{2}\alpha \left( {x-1} \right)$.
And ${\cal F}_{n,\nu }^C$ is proportional to $\mathbb{H} ={\rm{HeunC}}(\alpha ,\beta ,\gamma ,\delta ,\eta ;x)$, such that $ \mathbb{H} \sim{\cal F}_{n,\nu }^C$. The expression of ${\cal F}_{n,\nu }^C$ can written as
\begin{equation}\label{eq:HeunC-FC}
{\cal F}_{n,\nu }^C(\hat z) = {\left( {\frac{{ \alpha }}{2}} \right)^\tau }{{\rm{e}}^{ - \frac{{i\pi \tau }}{2} - \frac{\alpha }{2} + i\hat z}}{{\hat z}^{ - \frac{{\gamma  + \beta  + 2}}{2}}}f_{n,\nu }^C(\hat z),
\end{equation}
where $\tau  = \frac{1}{4}\left( {3\beta  + \gamma  + \frac{{2\delta }}{\alpha }} \right)$, and $f_{n,\nu }^C$ can be defined as
\begin{equation}\label{fnnc}
f_{n,\nu }^C = \mathbb{F} \sum\limits_{n =  - \infty }^\infty  {{{( - i)}^n}\frac{{{{(\nu  + 1 + i\hat \eta )}_n}}}{{{{(\nu  + 1 - i\hat \eta )}_n}}}f_n^\nu } {F_{n + \nu }}(\hat \eta ,\hat z),
\end{equation}
in which the Coulomb wave function ${F_{n + \nu }}(\hat \eta ,\hat z)$ can be defined as
\begin{equation}
{F_{n + \nu }}(\hat \eta ,\hat z) = {2^{n + \nu }}{{\hat z}^{n + \nu  + 1}}{e^{ - i\hat z}}\frac{{\Gamma (\hat a)}}{{\Gamma (\hat c)}}\Phi \left( {\hat a,\hat c;2i\hat z} \right),
\end{equation}
with $ \hat a = n + \nu  + 1 + \frac{{\delta }}{\alpha }$ , $\hat c = 2n + 2\nu  + 2,$  and $ i{\hat \eta}  =-\frac{{\delta }}{\alpha }$, $\Phi$ is the regular confluent hypergeometric function which can be represented by the following series expansion \cite{olver2010nist}:
\begin{equation}
  \Phi \left( {\hat a,\hat c;2i\hat z} \right) = \sum\limits_{j = 0}^\infty  {\frac{{{{(\hat a)}_j}}}{{{{(\hat c)}_j}}}{{\frac{{\left( {2i\hat z} \right)}}{{j!}}}^j}}.
\end{equation}
Expanding $f_{n,\nu }^C$ to series of $\hat{z}$
\begin{equation}\label{eq:fc-series}
f_{n,\nu }^C = \mathbb{F} {e^{ - i\hat z}}{2^\nu }{{\hat z}^{\nu  + 1}}\sum\limits_{k =  - \infty }^\infty  {\sum\limits_{n =  - \infty }^k {{D_{n,n - k}}} } {{\hat z}^k},
\end{equation}
where
\begin{align}
{D_{n,k - n}} &= \frac{{\Gamma \left( {n + \nu + 1 + \frac{\delta }{\alpha }} \right){{\left( {\nu  + 1 - \frac{\delta }{\alpha }} \right)}_n}}}{{\Gamma \left( {2n + 2\nu  + 2} \right){{\left( {\nu  + 1 + \frac{\delta }{\alpha }} \right)}_n}}}\nonumber\\
& \times \frac{{{{\left( {n + \nu  + 1 + \frac{\delta }{\alpha }} \right)}_{k - n}}}}{{{{(2n + 2\nu  + 2)}_{k - n}}(k - n)!}}{( - 1)^n}{(2i)^k}f_n^\nu .
\end{align}

Because $f_{n,\nu }^C ({\hat z})$ is convergent at infinity, then we discuss its analytic asymptotic formula at infinity.
There is the analytic property \cite{Figueiredo2002,El_Jaick_2013,bateman1953higher} of the confluent hypergeometric function
\begin{align}
\Phi (\hat a,\hat c;2i\hat z) &= \frac{{\Gamma (\hat c)}}{{\Gamma (\hat c - \hat a)}}{e^{i\epsilon \hat a\pi }}\Psi (\hat a,\hat c;2i\hat z)\\
 &+ \frac{{\Gamma (\hat c)}}{{\Gamma (\hat a)}}{e^{i\pi (\hat a - \hat c)\epsilon }}{e^{2i\hat z}}\Psi (\hat c - \hat a,\hat c; - 2i\hat z), \nonumber
\end{align}
with
\begin{equation}
  {\epsilon  = {\rm{sgn}}\left( {{\rm{Im}}(2i\hat z)} \right) = \left\{ {\begin{array}{*{20}{l}}
{1,\quad \quad {\rm{if}}\quad {\rm{Im}}(2i\hat z) > 0,}\\
{ - 1,\quad {\rm{if}}\quad {\rm{Im}}(2i\hat z) < 0,}
\end{array}} \right.}
\end{equation}
where  $\Psi$ is the irregular confluent hypergeometric function. Then, Eq. (\ref{fnnc}) can be expressed as
\begin{subequations}
  \begin{align}
&f_{n,\nu }^C =  f_{n,\nu }^ \odot +f_{n,\nu }^ \otimes \label{eq:fnC2} \\
&\begin{array}{l}
f_{n,\nu }^ \odot  ={2^\nu } \mathbb{F} {e^{i\pi \left( {\nu  + 1+ \frac{\delta }{\alpha }} \right)}}{{\hat z}^{\nu  + 1}}{e^{ - i\hat z}}\frac{{\Gamma \left( {\nu  + 1 + \frac{\delta }{\alpha }} \right)}}{{\Gamma \left( {\nu  + 1 - \frac{\delta }{\alpha }} \right)}}\\
\quad \quad  \times \sum\limits_{n =  - \infty }^\infty  {f_n^\nu {{(2i\hat z)}^n}} \Psi \left( {\hat a,\hat c,2i\hat z} \right),
\end{array}\\
&\begin{array}{l}
f_{n,\nu }^ \otimes  = {2^\nu }\mathbb{F}{e^{ - i\pi \left( {\nu  + 1 - \frac{\delta }{\alpha }} \right)}}{{\hat z}^{\nu  + 1}}{e^{i\hat z}}\sum\limits_{n =  - \infty }^\infty  {\frac{{{{\left( {\nu  + 1 - \frac{\delta }{\alpha }} \right)}_n}}}{{{{\left( {\nu  + 1 + \frac{\delta }{\alpha }} \right)}_n}}}} \\
\quad \quad  \times {( - 1)^n}f_n^\nu {( - 2i\hat z)^n}\Psi (\hat c - \hat a,\hat c; - 2i\hat z).
\end{array}
\end{align}
\end{subequations}

By taking into account the asymptotic behavior \cite{olver2010nist} of $\Psi\left( {\hat a,\hat c;2i\hat z} \right)$ at large $|x|$:
\begin{equation}\label{irrch}
  \mathop {\lim }\limits_{x \to \infty } \Psi\left( {\hat a,\hat c;2i\hat z} \right)\rightarrow{(2i\hat z)^{ - \hat{a}}},
\end{equation}
we can obtain the asymptotic analytic expression of $f_{n,\nu }^C $ at infinity:
\begin{subequations}
  \begin{align}
&f_{n,\nu }^ \odot  = A_{n,\nu }^ \odot {x^{ - \frac{\delta }{\alpha }}}{{\rm{e}}^{\frac{\alpha }{2}x}},\quad f_{n,\nu }^ \otimes  = A_{n,\nu }^ \otimes {x^{\frac{\delta }{\alpha }}}{{\rm{e}}^{ - \frac{\alpha }{2}x}},\\
&
A_{n,\nu }^ \odot  = {{\rm{e}}^{ - \frac{\alpha }{2}}}{\left( {\frac{{i\alpha }}{2}} \right)^{ - \frac{\delta }{\alpha }}}\tilde A_{n,\nu }^ \odot ,{\kern 1pt}
A_{n,\nu }^ \otimes  = {{\rm{e}}^{\frac{\alpha }{2}}}{\left( {\frac{{i\alpha }}{2}} \right)^{\frac{\delta }{\alpha }}}\tilde A_{n,\nu }^ \otimes .
\end{align}
\end{subequations}
where
\begin{subequations}
  \begin{align}
&\tilde A_{n,\nu }^ \odot  = {2^{ - 1 - \frac{\delta }{\alpha }}}{{\rm{e}}^{\frac{{i\pi }}{2}\left( {\nu  + 1 + \frac{\delta }{\alpha }} \right)}}\frac{{\Gamma \left( {\nu  + 1 + \frac{\delta }{\alpha }} \right)}}{{\Gamma \left( {\nu  + 1 - \frac{\delta }{\alpha }} \right)}},\\
&\begin{array}{l}
\tilde A_{n,\nu }^ \otimes  = {2^{ - 1 + \frac{\delta }{\alpha }}}{{\rm{e}}^{ - \frac{{i\pi }}{2}\left( {\nu  + 1 - \frac{\delta }{\alpha }} \right)}}{\left( {\sum\limits_{n =  - \infty }^{ + \infty } {f_n^\nu } } \right)^{ - 1}}\\
\quad \quad \times
\left( {\sum\limits_{n =  - \infty }^{ + \infty } {{{( - 1)}^n}} \frac{{{{\left( {\nu  + 1 - \frac{\delta }{\alpha }} \right)}_n}}}{{{{\left( {\nu  + 1 + \frac{\delta }{\alpha }} \right)}_n}}}f_n^\nu } \right).
\end{array}
\end{align}
\end{subequations}

Analogous to \cref{eq:fnC2}, we can derive the asymptotic analytical expression of $f_{-n,-\nu-1}^C$ as $|x| \rightarrow \infty$,
\begin{equation}
{f_{ - n, - \nu  - 1}^C = {{\rm{e}}^{i\pi \left( {\nu  + \frac{1}{2}} \right)}}f_{n,\nu }^ \otimes } + \frac{{\sin \pi \left( {\nu  + \frac{\delta }{\alpha }} \right)}}{{\sin \pi \left( {\nu  - \frac{\delta }{\alpha }} \right)}}{{\rm{e}}^{ - i\pi \left( {\nu  + \frac{1}{2}} \right)}}f_{n,\nu }^ \odot  .
\end{equation}

\subsection{Proportionality Coefficient}\label{sec:PC}
Both solutions, ${\mathbb{H} _{n,\nu }}$ and ${{\cal F}_{n,\nu }^C}$, converge within an extremely wide region of $\hat z$.
As $k$ is an arbitrary integer, we set $k=0$, and find that the series representations ( \Cref{eq:hyper-expand-2,eq:HeunC-FC,eq:fc-series} ) of ${\mathbb{H} _{n,\nu }}$ and ${{\cal F}_{n,\nu }^C}$ are proportional to the same single-valued function of $\hat z$.
Therefore, the analytical properties of ${\mathbb{H} _{n,\nu }}$ and ${{\cal F}_{n,\nu }^C}$ are identical, indicating that these two solutions are equivalent up to a multiplicative constant.
The proportional coefficient between ${{\mathbb{H} _{n,\nu }}}$ and ${{{\cal F}_{n,\nu }^C }}$ can now be determined
\begin{equation}\label{eq:Xi1}
\Xi _{n,\nu }^\beta  = \frac{{{\mathbb{H} _{n,\nu }}}}{{{\cal F}_{n,\nu }^C}} = {2^{ - \nu }}{\left( {\frac{\alpha }{2}} \right)^{ - \hat \tau }}{{\rm{e}}^{\frac{{i\pi \hat \tau  + \alpha }}{2}}}\frac{{\sum\limits_{j = 0}^\infty  {{C_{n,n - k}}} }}{{\sum\limits_{n =  - \infty }^k {{D_{n,n - k}}} }},
\end{equation}
where $\hat \tau  = \frac{{\beta  - \gamma }}{4} + \nu  + \frac{\delta }{{2\alpha }}.$

For the convenience of calculation, we set $k = 0$, so Eq. \eqref{eq:Xi1} is simplified as
\begin{align}\label{eq:Xinv}
&\begin{array}{*{20}{l}}
{\Xi _{n,\nu }^\beta  = \frac{{{2^{ - \nu }}{{\left( {\frac{\alpha }{2}} \right)}^{ - \hat \tau }}{{\rm{e}}^{\frac{{i\pi \hat \tau  + \alpha }}{2}}}\Gamma \left( {\beta  + 1} \right)\Gamma \left( {2\nu  + 2} \right)}}{{\Gamma \left( {\nu  + 1 + \frac{\delta }{\alpha }} \right)\Gamma \left( {\nu  + 1 - \frac{{\beta  + \gamma }}{2}} \right)\Gamma \left( {\nu  + 1 + \frac{{\gamma  - \beta }}{2}} \right)}}}\\
{ \times {{\Big( {\sum\limits_{n =  - \infty }^0 {\frac{{{{( - 1)}^n}{{\left( {\nu  + 1 - \frac{\delta }{\alpha }} \right)}_n}}}{{( - n)!{{\left( {2\nu  + 2} \right)}_n}{{\left( {\nu  + 1 + \frac{\delta }{\alpha }} \right)}_n}}}f_n^\nu } } \Big)}^{ - 1}}}
\end{array}\\
&\begin{array}{l}
{ \times \Big( {\sum\limits_{n = 0}^\infty  {{{\left( { - 1} \right)}^n}\frac{{\Gamma \left( {n + 2\nu  + 1} \right)\Gamma \left( {n + \nu  + 1 + \frac{{\gamma  - \beta }}{2}} \right)\Gamma \left( {n + \nu  + 1 - \frac{{\beta  + \gamma }}{2}} \right)}}{{( n!)\Gamma \left( {n + \nu  + 1 + \frac{{\beta  - \gamma }}{2}} \right)\Gamma \left( {n + \nu  + 1 + \frac{{\beta  + \gamma }}{2}} \right)}}f_n^\nu } } \Big)}.
\end{array}\nonumber
\end{align}
To obtain the proportional coefficients $\Xi _{ - n, - \nu  - 1}^{\beta}$ for ${\mathbb{H} _{-n,-\nu-1 }}$ and ${{\cal F}_{-n,-\nu-1 }^C}$, we can simply substitute $n\Rightarrow-n$ and $\nu\Rightarrow-\nu-1$  into \cref{eq:Xinv}.

\subsection{Infinite Asymptotic Behavior of HeunC Function}
Based on the previous sections, we are now able to derive the analytical asymptotic expression for the confluence Heun function at infinity. Specifically, we can rewrite \cref{eq:hc-inf} as follows.

\begin{align}\label{eq:87}
\mathop {\lim }\limits_{|x| \to \infty } \mathbb{H} (x) &= \Xi _{n,\nu }^\beta {\cal F}_{n,\nu }^C + \Xi _{ - n, - \nu  - 1}^\beta {\cal F}_{ - n, - \nu  - 1}^C\nonumber\\
 &= D_ \odot^\beta \;{x^{ - \frac{{\beta  + \gamma  + 2}}{2} - \frac{\delta }{\alpha }}} + D_\otimes ^\beta {{\rm{e}}^{ - \alpha x}}{x^{ - \frac{{\beta  + \gamma  + 2}}{2} + \frac{\delta }{\alpha }}} .
\end{align}

Substituting the results from \cref{sec:Expan-S-CWF,sec:PC} into \cref{eq:87}, the constants $D_ \odot ^\beta$ and $D_ \otimes ^\beta$ in the asymptotic behavior at infinity (\ref{eq:hc-inf}) are given by
\begin{align}
{}&\begin{array}{l}\label{eq:Dinc}
D_ \odot ^\beta  = \Xi _{n,\nu }^\beta D_{ \odot ,n,\nu }^\beta \\
 \quad \quad+ {{\rm{e}}^{ - i\pi \left( {\nu  + \frac{1}{2}} \right)}}\frac{{\sin \pi \left( {\nu  + \frac{\delta }{\alpha }} \right)}}{{\sin \pi \left( {\nu  - \frac{\delta }{\alpha }} \right)}}\Xi _{ - n, - \nu  - 1}^\beta D_{ \odot , - n, - \nu  - 1}^\beta ,
\end{array}\\
{}&{D_ \otimes ^\beta  = \Xi _{n,\nu }^\beta D_{ \otimes ,n,\nu }^\beta  + {{\rm{e}}^{i\pi \left( {\nu  + \frac{1}{2}} \right)}}\Xi _{ - n, - \nu  - 1}^\beta D_{ \otimes , - n, - \nu  - 1}^\beta ,}\label{eq:Dout}
\end{align}
with
\begin{align}
D_{ \odot ,n,\nu }^\beta  &= {\left( { - 1} \right)^{\frac{{\gamma  + \beta  + 2}}{2} + \frac{\delta }{\alpha }}}{\big( {\frac{\alpha }{2}} \big)^\tau } {\big( -{\frac{{i\alpha }}{2}} \big)^{ - \frac{{\gamma  + \beta  + 2}}{2} - \frac{\delta }{\alpha }}}{{\rm{e}}^{ - \frac{{i\pi \tau +\alpha}}{2}}}
\nonumber \\
&\times
{2^{ - 1 - \frac{\delta }{\alpha }}}{{\rm{e}}^{\frac{{i\pi }}{2}\big( {\nu  + 1 + \frac{\delta }{\alpha }} \big)}} \Xi _{n,\nu }^\beta\frac{{\Gamma \big( {\nu  + 1 + \frac{\delta }{\alpha }} \big)}}{{\Gamma \big( {\nu  + 1 - \frac{\delta }{\alpha }} \big)}}, \\
 D_{  \otimes,n,\nu }^\beta  &= {\big( { - 1} \big)^{\frac{{\gamma  + \beta  + 2}}{2} - \frac{\delta }{\alpha }}}{\left( {\frac{\alpha }{2}} \right)^\tau } {\left(- {\frac{{i\alpha }}{2}} \right)^{ - \frac{{\gamma  + \beta  + 2}}{2} + \frac{\delta }{\alpha }}}\nonumber \\
&\times  {{\rm{e}}^{ - \frac{{i\pi \tau - \alpha}}{2}}}\Xi _{n,\nu }^\beta  \frac{ {2^{ - 1 + \frac{\delta }{\alpha }}}{{\rm{e}}^{ - \frac{{i\pi }}{2}\big( {\nu  + 1 - \frac{\delta }{\alpha }} \big)}}} {\sum\limits_{n =  - \infty }^{ + \infty } {f_n^\nu } }   \nonumber \\
&\times\Big( {\sum\limits_{n =  - \infty }^{ + \infty } {{{( - 1)}^n}} \frac{{{{\big( {\nu  + 1 - \frac{\delta }{\alpha }} \big)}_n}}}{{{{\big( {\nu  + 1 + \frac{\delta }{\alpha }} \big)}_n}}}f_n^\nu } \Big).
\end{align}

\section{Application to Schwarzschild BHs}\label{sec:Application}
For simplicity but without loss of generality, we shall consider the gravitational, electromagnetic, and scalar fluxes of the Schwarzschild black hole as an illustrative example, and we shall provide the complete solution for the purely ingoing wave at the horizon and the purely outgoing wave at infinity, along with their respective amplitudes. We consider the case when a test particle of mass $\mu$ travels a circular orbit around a Schwarzschild BH of mass $M\gg \mu$. To calculate the radiation fluxes, the radial Teukolsky equation of a Schwarzschild BH with the point source was given by \cref{eq:HRTEs}, reduces to:

\begin{align}\label{eq:Steuk}
&\Delta R_{\ell m\omega }^{\prime \prime } + 2(r - M)(s + 1)R_{\ell m\omega }^\prime \\
 &+ \left[ {r^2}\left( {{\omega ^2}{r^2} - 4i\omega (r - 3M)} \right) - \Delta  {\lambda} \right]{R_{\ell m\omega }} = {}_s T_{\ell m\omega}, \nonumber
\end{align}
where $\Delta  = r( r - {r_{\rm H}} )$, and ${r_{\rm H}}=2M$. The source term ${}_s T_{\ell m\omega}$ for the gravitational, electromagnetic, and scalar perturbations are given in Ref. \cite{Teukolsky1973}.

Using Green's function method, it is necessary to construct  two linear independent solutions of \cref{eq:Steuk}, denoted as $R^{\rm in}_{\ell m\omega}(r)$ and $R^{\rm up}_{\ell m\omega}(r)$, that satisfy the following boundary conditions:
\begin{align}
    & R_{\ell m\omega }^{{\rm{in}}}\to \left\{ {
    \begin{array}{*{20}{l}}
{B_{\ell m\omega }^{{\rm{trans}}}{\Delta ^{ - s}}{{\rm{e}}^{ - i\omega{r^*}}},}&{r \to r_{\rm H}},\\
{{{B_{\ell m\omega }^{{\rm{ref}}}}}{r^{1-2s}}{{\rm{e}}^{i\omega {r^*}}} + {{B_{\ell m\omega }^{{\rm{inc}}}}}{r^{-1}}{{\rm{e}}^{ - i\omega {r^*}}},}&{r \to  + \infty ,}
\end{array}
    } \right.\label{eq:boundary1}\\
 & R_{\ell m\omega }^{{\rm{up}}} \to \left\{ {\begin{array}{*{20}{l}}
{C_{\ell m\omega }^{{\rm{up}}}{{\rm{e}}^{i\omega{r^*}}} + {C_{\ell m\omega }^{{\rm{ref}}}}{\Delta ^{-s}}{{\rm{e}}^{ - i\omega{r^*}}},}&{r \to {r_{\rm H} },}\\
{{{C_{\ell m\omega }^{{\rm{trans}}}}}{r^{1-2s}}{{\rm{e}}^{i\omega {r^*}}},}&{r \to  + \infty ,}
\end{array}} \right.\label{eq:boundary2}
\end{align}
where
$ r^{*}=r+r_{\rm H} \ln(r/r_{\rm H}-1)$
and $s=-2$.

In the case of a circular orbit, the specific energy $\tilde E$ and angular momentum $\tilde L_z$ of the particle are given by
\begin{align}
    &\tilde E =(r_0-2M)/\sqrt{r_0(r_0-3M)}, \label{eq:ene}\\
 & \tilde L_z =\sqrt{Mr_0}/\sqrt{1-3M/r_0},
\end{align}
where $r_0$ is the orbital radius. The angular frequency is given by $\Omega=(M/r_0^3)^{1/2}$ , so the orbital frequency is $\omega = m \Omega$ and the orbital frequency in Mino time is $\Upsilon_t =({\frac{r_0^5}{r_0-3M}})^{1/2}$.

\subsection{Normalized solutions}

The solutions $R_{\ell m\omega }^{{\rm{in,up}}}$ proposed in \Cref{sec:in-outgongsol} are the analytical solutions satisfying the boundary conditions \eqref{eq:boundary1} and \eqref{eq:boundary2} of the Schwarzschild BH.
To facilitate comparisons with other methods, we need normalize $R_{\ell m\omega }^{{\rm{in}}}$ at the event horizon and $R_{\ell m\omega }^{{\rm{up}}}$ at infinity, obtaining normalized asymptotic amplitudes of homogeneous equations \eqref{eq:HRTEs}.
Using the normalized condition $B_{\ell m\omega }^{{\rm{trans}}} = 1$ at event horizon, the ingoing wave solution \eqref{eq:uHor} is normalized to obtain
\begin{equation}
\tilde R_{\ell m\omega }^{{\rm{in}}} =  S_{\rm H}^\beta \left( x \right){\rm{HeunC}}\left( {\alpha ,\beta ,\gamma ,\delta ,\eta ;x} \right),
\end{equation}
where the normalized S-homotopic transformation at horizon is
\begin{equation}
 S_{\rm H}^\beta \left( x \right) =r^{-2s}_{\rm{H}} {{\rm{e}}^{\frac{1}{2}\alpha \left( {x - 1} \right)}}{{{\left( { - x} \right)}^{\frac{1}{2}\left( {\beta  - s} \right)}}{{\left( {1 - x} \right)}^{\frac{1}{2}\left( {\gamma  - s} \right)}}}.
\end{equation}

By utilizing the asymptotic behavior \eqref{eq:boundary1} of the solution $\tilde R_{\ell m\omega }^{{\rm{in }}}$ as $r\rightarrow \infty$, we derive analytic expressions of the asymptotic amplitudes $B_{\ell m\omega }^{\rm inc}$ and $B_{\ell m\omega }^{\rm ref}$ for $\tilde R_{\ell m\omega }^{{\rm{in}}}$.
\begin{subequations}
  \begin{align}
  B_{\ell m\omega }^{{\rm{inc}}} &=  - r_{\rm{H}}^5D_ \odot ^\beta,\\
   B_{\ell m\omega }^{{\rm{ref}}} &= {\left( { - 1} \right)^{2i\omega {r_{\rm{H}}} - 1}}{r_{\rm{H}}}{{\rm{e}}^{ - 2i\omega {r_{\rm{H}}}}}D_ \otimes ^\beta .
\end{align}
\end{subequations}

Meanwhile, using the normalized condition $C_{\ell m\omega }^{{\rm{trans}}} = 1$ at infinity, the outgoing wave solution is normalized to obtain
\begin{align}
\tilde R_{\ell m\omega }^{{\rm{up}}} = & {\left( { - 1} \right)^{\beta+1} } \frac{{D_ \odot ^{ - \beta }}}{{D_ \odot ^\beta }}\tilde D S_\infty ^\beta (x){\rm{HeunC}}(\alpha ,\beta ,\gamma ,\delta ,\eta ;x) \nonumber \\
&+ \tilde D S_\infty ^{ - \beta }(x){\rm{HeunC}}(\alpha , - \beta ,\gamma ,\delta ,\eta ;x),
\end{align}
where the normalized S-homotopic transformation at infinity is
\begin{equation}
S_\infty ^\beta (x) =  - r_{\rm{H}}^3{{\rm{e}}^{\frac{1}{2}\alpha \left( {x + 1} \right)}}{\left( { - x} \right)^{\frac{1}{2}(\beta  - s)}}{\left( {1 - x} \right)^{\frac{1}{2}(\gamma  - s)}},
\end{equation}
and
\[\tilde D = {\left( {D_ \otimes ^{ - \beta } - \frac{{D_ \odot ^{ - \beta }}}{{D_ \odot ^\beta }}D_ \otimes ^\beta } \right)^{ - 1}}.\]

By utilizing the asymptotic behavior \eqref{eq:boundary2} of the solution $\tilde R_{\ell m\omega }^{{\rm{up }}}$ as $r\rightarrow r_{\rm{H}}$, we derive analytic expressions of the asymptotic amplitudes $C_{\ell m\omega }^{{\rm{ref}}}$ and $C_{\ell m\omega }^{{\rm{up}}}$ for $\tilde R_{\ell m\omega }^{{\rm{up}}} $.
\begin{subequations}
  \begin{align}
   C_{\ell m\omega }^{{\rm{up}}} &=  - {r}^3_{\rm{H}}\tilde D,\\
   C_{\ell m\omega }^{{\rm{ref}}} &= \frac{{{{\left( { - 1} \right)}^{ - 2i\omega {r_{\rm{H}}}}}}}{{{r_{\rm{H}}}}}\frac{{D_ \odot ^{ - \beta }}}{{\;D_ \odot ^\beta }}{{\rm{e}}^{2i\omega {r_{\rm{H}}}}}\tilde D.
\end{align}
\end{subequations}

For the ingoing wave solution $\tilde R_{\ell m\omega }^{{\rm{in }}}$ and outgoing wave solution $\tilde R_{\ell m\omega }^{{\rm{up }}}$, the parameters ($\alpha,\beta,\gamma,\delta$, $\eta$) of these two solutions for the Schwarzschild BH can be seen in \Cref{table:parameter}.

\subsection{Energy Fluxes of Schwarzschild BHs}

\subsubsection{Gravitational field}

According to BH perturbation theory, gravitational waves are described by perturbations in the Weyl scalars $\psi_0$ and $\psi_4$. For vacuum solutions corresponding to pure gravitational waves, both $\psi_0$ and $\psi_4$ contain the same information about the propagation of gravitational waves within the Kerr spacetime.
However, both for near null infinity and near the horizon, $\psi_4$ exhibits dominance over $\psi_0$ for gravitational waves propagating along the positive radial ($+r$) direction, whereas $\psi_0$ prevails over $\psi_4$ for waves propagating along the negative radial ($-r$) direction.
Specifically, $\psi_0$ corresponds to gravitational waves with spin-weight $s = +2$, while $\psi_4$ corresponds to gravitational waves with spin-weight $s = -2$.
In studies focusing on gravitational waves generated by compact binary coalescence that propagate towards future null infinity, it is natural to compute $\psi_4$. Thus, Teukolsky equations have been extensively applied for accurate evaluations of $\psi_4$ in the pertinent scientific literature.

In order to calculate the radiative energy and angular momentum flues, the expression of the asymptotic amplitude ${}_s \tilde Z_{\ell m\omega}^{{\rm H},\infty}$ in inhomogeneous solution \eqref{eq:Rlmw-source}  needs to be solved.
The expressions of ${}_s \tilde Z_{\ell m\omega}^{{\rm H},\infty}$, ${_s{\beta _{\ell m\omega }}}$, and ${{}_s{\alpha _{\ell m\omega }}}$ can be found in \Cref{app:expression}.
Finally, theses coefficients and functions are substituted into \Cref{eq:dEdtInf,eq:dEdtH}, and the expressions of energy fluxes are obtained.
Therefore, we obtain energy fluxes of the gravitational perturbation field of the spin-weight $s = -2$.
\begin{subequations}
  \begin{align}
 {\textcolor{white}{\biggl{|}}}_{ - 2} \left\langle {\frac{{dE}}{{dt}}} \right\rangle _\infty&= \sum\limits_{\ell  = 2}^\infty  {\sum\limits_{m =  - \ell }^\ell  {\frac{{16\pi }}{{{\omega ^2}}}\left| {\frac{1}{{{W_{\rm{C}}}{\Upsilon _t}}}\left[ {\left( {{A_{nn0}} + {A_{\bar mn0}} + {A_{\bar m\bar m0}}} \right)R_{\ell m\omega }^{{\rm{in}}}} \right.} \right.} } \nonumber \\
&{\left. { - \left( {{A_{\bar mn1}} + {A_{\bar m\bar m1}}} \right){{\left( {R_{\ell m\omega }^{{\rm{in}}}} \right)}^\prime }\left. { + {A_{\bar m\bar m2}}{{\left( {R_{\ell m\omega }^{{\rm{in}}}} \right)}^{\prime \prime }}} \right]} \right|^2},
\end{align}
\begin{align}
 {\textcolor{white}{\biggl{|}}}_{-2} \left\langle {\frac{{dE}}{{dt}}} \right\rangle _{\rm{H}}&= \sum\limits_{\ell  = 2}^\infty  {\sum\limits_{m =  - \ell }^\ell  {\frac{{{\rm{4096}}\pi {{(2M{r_{\rm{H}}})}^5}({\omega ^2} + 4{{\tilde \epsilon }^2})({\omega ^2} + 16{{\tilde \epsilon }^2}){\omega ^2}}}{{{\lambda ^2}{{(\lambda  + 2)}^2} + 144{\omega ^2}{M^2}.}}} } \nonumber\\
& \times \left| {\frac{1}{{{W_{\rm{C}}}{\Upsilon _t}}}\left[ {\left( {{A_{nn0}} + {A_{\bar mn0}} + {A_{\bar m\bar m0}}} \right)R_{\ell m\omega }^{{\rm{up}}}} \right.} \right.\nonumber\\
&\left. { - \left( {{A_{\bar mn1}} + {A_{\bar m\bar m1}}} \right){{\left( {R_{\ell m\omega }^{{\rm{up}}}} \right)}^\prime }\left. { + {A_{\bar m\bar m2}}{{\left( {R_{\ell m\omega }^{{\rm{up}}}} \right)}^{\prime \prime }}} \right]} \right|^2.
\end{align}
\end{subequations}
where $\tilde \epsilon  = {(4{r_{\rm{H}}})^{ - 1}}$ and the explicit form of $A_{nn0}$ and other terms can be found and derived in Ref. \cite{Sasaki_2003,Teukolsky1973}.

Similarly, substituting ${}_s \tilde Z_{\ell m\omega}^{{\rm H},\infty}$, ${_s{\beta _{\ell m\omega }}}$, and ${{}_s{\alpha _{\ell m\omega }}}$ into the radiative energy fluxes \eqref{eq:dEdtInf} and \eqref{eq:dEdtH}, we obtain energy fluxes of the gravitational perturbation field of the spin-weight $s = +2$.
\begin{subequations}
  \begin{align}
  {\textcolor{white}{\biggl{|}}}_{ + 2}\left\langle {\frac{{dE}}{{dt}}} \right\rangle _\infty &= \sum\limits_{\ell  = 2}^\infty  {\sum\limits_{m =  - \ell }^\ell  {\frac{{{\rm{256}}\pi {\omega ^6}}}{{{{(\lambda  + 4)}^2}{{(\lambda  + 6)}^2} + 144{M^2}{\omega ^2}}}} }  \nonumber \\
 &\times \left| {\frac{1}{{{W_{\rm{C}}}{\Upsilon _t}}}\left[ {\left( {{A_{ll0}} + {A_{lm0}} + {A_{mm0}}} \right)R_{\ell m\omega }^{{\rm{in}}}} \right.} \right.  \nonumber \\
&{\left. { - \left( {{A_{lm1}} + {A_{mm1}}} \right){{\left( {R_{\ell m\omega }^{{\rm{in}}}} \right)}^\prime }\left. { + {A_{mm2}}{{\left( {R_{\ell m\omega }^{{\rm{in}}}} \right)}^{\prime \prime }}} \right]} \right|^2},
\end{align}
\begin{align}
 {\textcolor{white}{\biggl{|}}}_{ + 2}\left\langle {\frac{{dE}}{{dt}}} \right\rangle _{\rm{H}} &= \sum\limits_{\ell  = 2}^\infty  {\sum\limits_{m =  - \ell }^\ell  {\frac{\pi }{{8r_{\rm{H}}^3\left( {{\omega ^2} + 4{\epsilon ^2}} \right)}}} } \nonumber\\
& \times \left| {\frac{1}{{{W_{\rm{C}}}{\Upsilon _t}}}\left[ {\left( {{A_{ll0}} + {A_{lm0}} + {A_{mm0}}} \right)R_{\ell m\omega }^{{\rm{up}}}} \right.} \right. \nonumber\\
&{\left. { - \left( {{A_{lm1}} + {A_{mm1}}} \right){{\left( {R_{\ell m\omega }^{{\rm{up}}}} \right)}^\prime }\left. { + {A_{mm2}}{{\left( {R_{\ell m\omega }^{{\rm{up}}}} \right)}^{\prime \prime }}} \right]} \right|^2}.
\end{align}
\end{subequations}

\subsubsection{Electromagnetic field}
Electromagnetic waves are described in terms of perturbations in the electromagnetic scalars $\phi_0$ and $\phi_2$.
The electromagnetic scalars $\phi_0$ and $\phi_2$, similar to Weyl scalars $\psi_0$ and $\psi_4$, both contain the same information of wave propagation in Kerr spacetime. However, when the electromagnetic wave propagates along the positive radial ($+r$) direction, $\phi_2$ dominates ; when the electromagnetic wave propagates along the negative radial ($-r$) direction, $\phi_0$ dominates.
Here, $\phi_0$ corresponds to the electromagnetic field with $s = +1$, and $\phi_2$ corresponds to the electromagnetic perturbation field with $s = -1$.

Similar to energy fluxes of GW radiation, ${}_s \tilde Z_{\ell m\omega}^{{\rm H},\infty}$, ${_s{\beta _{\ell m\omega }}}$, and ${{}_s{\alpha _{\ell m\omega }}}$ are substituted into the radiative energy fluxes \eqref{eq:dEdtInf} and \eqref{eq:dEdtH}, we obtain energy fluxes of the electromagnetic perturbation field of the spin-weight $s = -1$.
\begin{subequations}
  \begin{equation}
  {\textcolor{white}{\biggl{|}}}_{ - 1} \left\langle {\frac{{dE}}{{dt}}} \right\rangle _\infty = \sum\limits_{\ell  = 2}^\infty  {\sum\limits_{m =  - \ell }^\ell  {\frac{1}{{2\pi }}{{\left| {\frac{\mathfrak{A}_{-}}{{{W_{\rm{C}}}\Delta }}R_{\ell m\omega }^{{\rm{in}}} - \frac{\mathfrak{B}_{-}}{{{W_{\rm{C}}}\Delta }}{{\left( {R_{\ell m\omega }^{{\rm{in}}}} \right)}^\prime }} \right|}^2}} } ,
  \end{equation}
\begin{align}
 {\textcolor{white}{\biggl{|}}}_{ - 1}\left\langle {\frac{{dE}}{{dt}}} \right\rangle _{\rm{H}} &= \sum\limits_{\ell  = 2}^\infty  {\sum\limits_{m =  - \ell }^\ell  {\frac{{16{r_{\rm{H}}}{M^3}{\omega ^2}\left( {4{r_{\rm{H}}}^2{\omega ^2} + 1} \right)}}{{\pi {\lambda ^2}}}} } \nonumber \\
& \times {\left| {\frac{\mathfrak{A}_{-}}{{{W_{\rm{C}}}\Delta }}R_{\ell m\omega }^{{\rm{up}}} - \frac{\mathfrak{B}_{-}}{{{W_{\rm{C}}}\Delta }}{{\left( {R_{\ell m\omega }^{{\rm{up}}}} \right)}^\prime }} \right|^2}.
\end{align}
\end{subequations}
where the explicit form of ${\mathfrak{A}_{\pm}}$ and ${\mathfrak{B}_{\pm}}$ can be found in Eq. (48) of Ref. \cite{Torres_2022}

Similarly, substituting ${}_s \tilde Z_{\ell m\omega}^{{\rm H},\infty}$, ${_s{\beta _{\ell m\omega }}}$, and ${{}_s{\alpha _{\ell m\omega }}}$ into the radiative energy fluxes \eqref{eq:dEdtInf} and \eqref{eq:dEdtH}, we obtain energy fluxes of the electromagnetic perturbation field of the spin-weight $s = +1$.
\begin{subequations}
  \begin{equation}
 {\textcolor{white}{\biggl{|}}}_{ + 1} \left\langle {\frac{{dE}}{{dt}}} \right\rangle _\infty= {\sum\limits_{\ell  = 2}^\infty  {\sum\limits_{m =  - \ell }^\ell  {\frac{{2{\omega ^4}}}{{\pi {{(\lambda  + 2)}^2}}}\left| {\frac{\mathfrak{A}_{+}}{{{W_{\rm{C}}}}}R_{\ell m\omega }^{{\rm{in}}} - \frac{ \mathfrak{B}_{+}}{{{W_{\rm{C}}}\Delta }}{{\left( {\Delta R_{\ell m\omega }^{{\rm{in}}}} \right)}^\prime }} \right|} } ^2},
  \end{equation}
\begin{equation}
  {\textcolor{white}{\biggl{|}}}_{ + 1} \left\langle {\frac{{dE}}{{dt}}} \right\rangle _{\rm{H}}= \sum\limits_{\ell  = 2}^\infty  {\sum\limits_{m =  - \ell }^\ell  {\frac{1}{{16\pi r_{\rm H} }}{{\left| {\frac{\mathfrak{A}_{+}}{{{W_{\rm{C}}}}}R_{\ell m\omega }^{{\rm{up}}} - \frac{ \mathfrak{B}_{+}}{{{W_{\rm{C}}}\Delta }}{{\left( {\Delta R_{\ell m\omega }^{{\rm{up}}}} \right)}^\prime }} \right|}^2}} } .
\end{equation}
\end{subequations}

\subsubsection{Scalar field}
Substituting ${}_s \tilde Z_{\ell m\omega}^{{\rm H},\infty}$, ${_s{\beta _{\ell m\omega }}}$, and ${{}_s{\alpha _{\ell m\omega }}}$ into the radiative energy fluxes \eqref{eq:dEdtInf} and \eqref{eq:dEdtH}, we obtain energy fluxes of the scalar perturbation field of the spin-weight $s = 0$.
\begin{subequations}
  \begin{equation}\label{eq:dEdtInfS0}
 {\textcolor{white}{\biggl{|}}}_0 \left\langle {\frac{{dE}}{{dt}}} \right\rangle _\infty = \sum\limits_{\ell  = 2}^\infty  {\sum\limits_{m =  - \ell }^\ell  {4\pi {\omega ^2}{{\left| {\frac{{r_0^2}}{{{W_{\rm{C}}}{\Upsilon _t}}}{{\kern 1pt} _0}{Y_{\ell m\omega }}R_{\ell m\omega }^{{\rm{in}}}} \right|}^2}} } ,
  \end{equation}
\begin{equation}\label{eq:dEdtHorS0}
 {\textcolor{white}{\biggl{|}}}_0 \left\langle {\frac{{dE}}{{dt}}} \right\rangle _{\rm{H}} = \sum\limits_{\ell  = 2}^\infty  {\sum\limits_{m =  - \ell }^\ell  {8\pi r_{\rm H}{\omega ^2}{{\left| {\frac{{r_0^2}}{{{W_{\rm{C}}}{\Upsilon _t}}}{{\kern 1pt} _0}{Y_{\ell m\omega }}R_{\ell m\omega }^{{\rm{up}}}} \right|}^2}} } .
\end{equation}
\end{subequations}
where ${}_{s}Y_{\ell m\omega }$ is the spin-weighted spherical harmonics.

The significant application of our method is to calculate radiation fluxes in which both $R_{\ell m\omega }^{{\rm{in}}}$  and $R_{\ell m\omega }^{{\rm{up}}}$ can be obtained using Eq. (\ref{eq:D2GSol1}) directly with the boundary conditions \eqref{eq:boundary1} and \eqref{eq:boundary2}.
Our method differs from traditional results obtained through post-Newtonian expansion and post-Minkowskian expansion, as it provides complete results without requiring series expansion. Theoretically, our method should be more efficient and accurate than the expansion method, and it has a broader scope of application as it is not limited by physical constraints such as low frequency, slow-motion or weak-field limits.

%

\section{Comparisons with other methods}\label{sec:results}

\begin{figure*}[htbp]
	\centering
\includegraphics[width=7in]{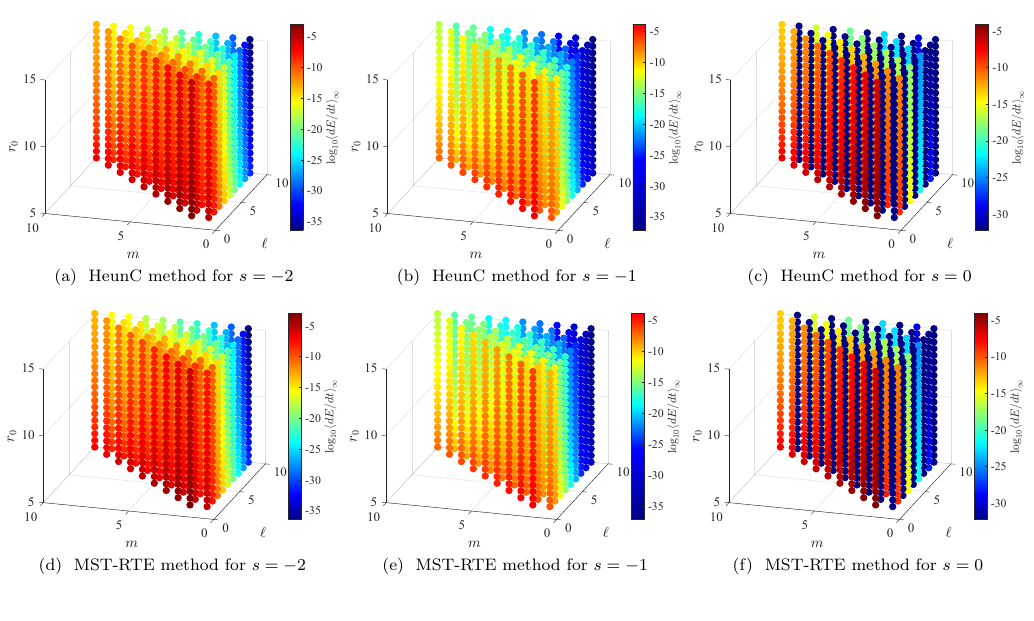}
\caption{The 3D figures of logarithmic luminosity at infinity  with  ${\bf N} = 50$, over the domain $[5M,15M] $. }\label{fig:FluxInf}
\end{figure*}
%
\begin{figure*}[htbp]
	\centering
\includegraphics[width=7in]{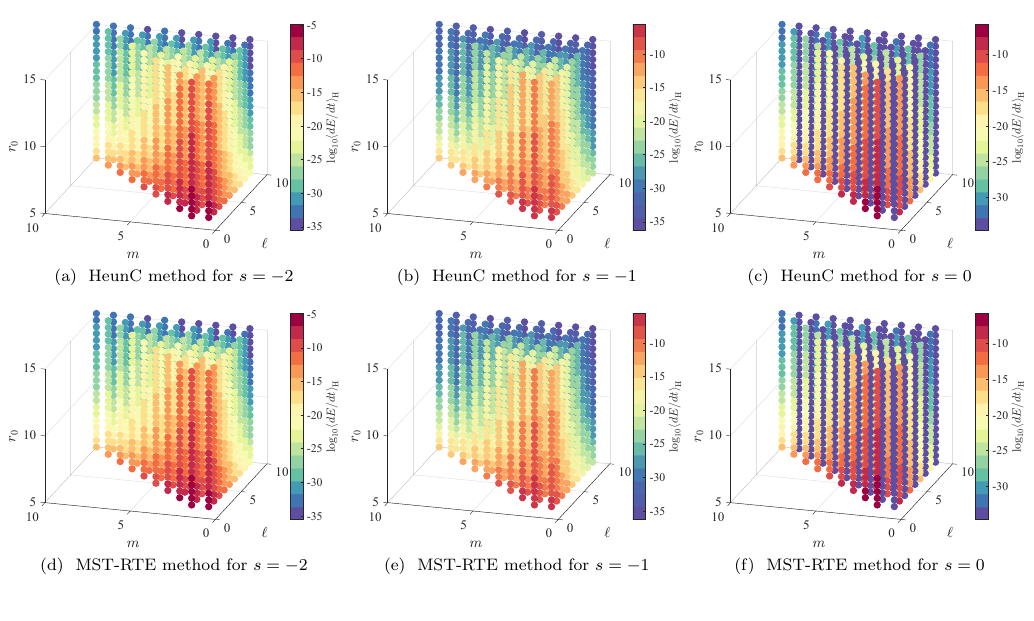}
\caption{The 3D figures of logarithmic luminosity at horizon  with  ${\bf N} = 50$, over the domain $[5M,15M] $. }\label{fig:FluxHor}
\end{figure*}

\begin{figure*}[htbp]
	\centering
\includegraphics[width=7in]{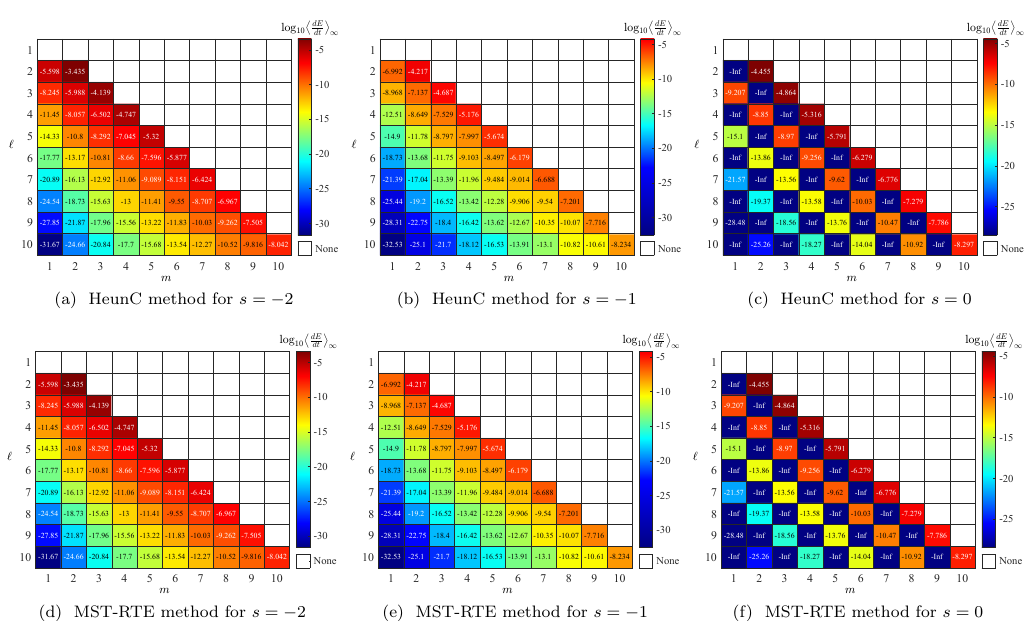}
\caption{The 2D figures of logarithmic luminosity at infinity  with  ${\bf N} = 50$ at $r_{\rm ISCO}$. }\label{fig:Flux2Inf}
\end{figure*}
\begin{figure*}[htbp]
	\centering
\includegraphics[width=7in]{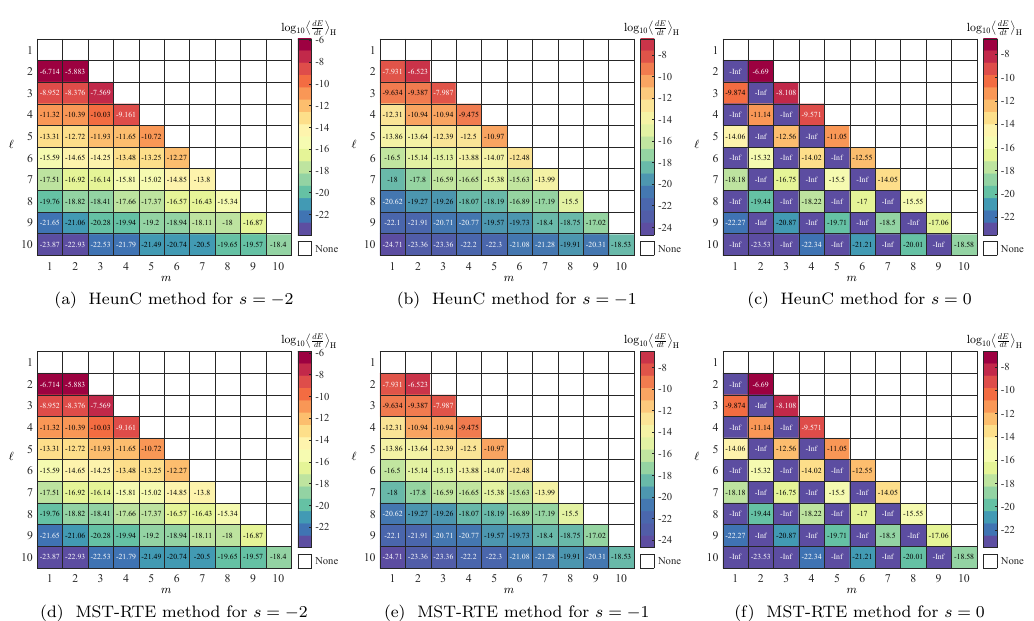}
\caption{The 2D figures of logarithmic luminosity at horizon  with  ${\bf N} = 50$ at $r_{\rm ISCO}$. }\label{fig:Flux2Hor}
\end{figure*}

\begin{figure*}[htbp]
	\centering
{\includegraphics[width=7in]{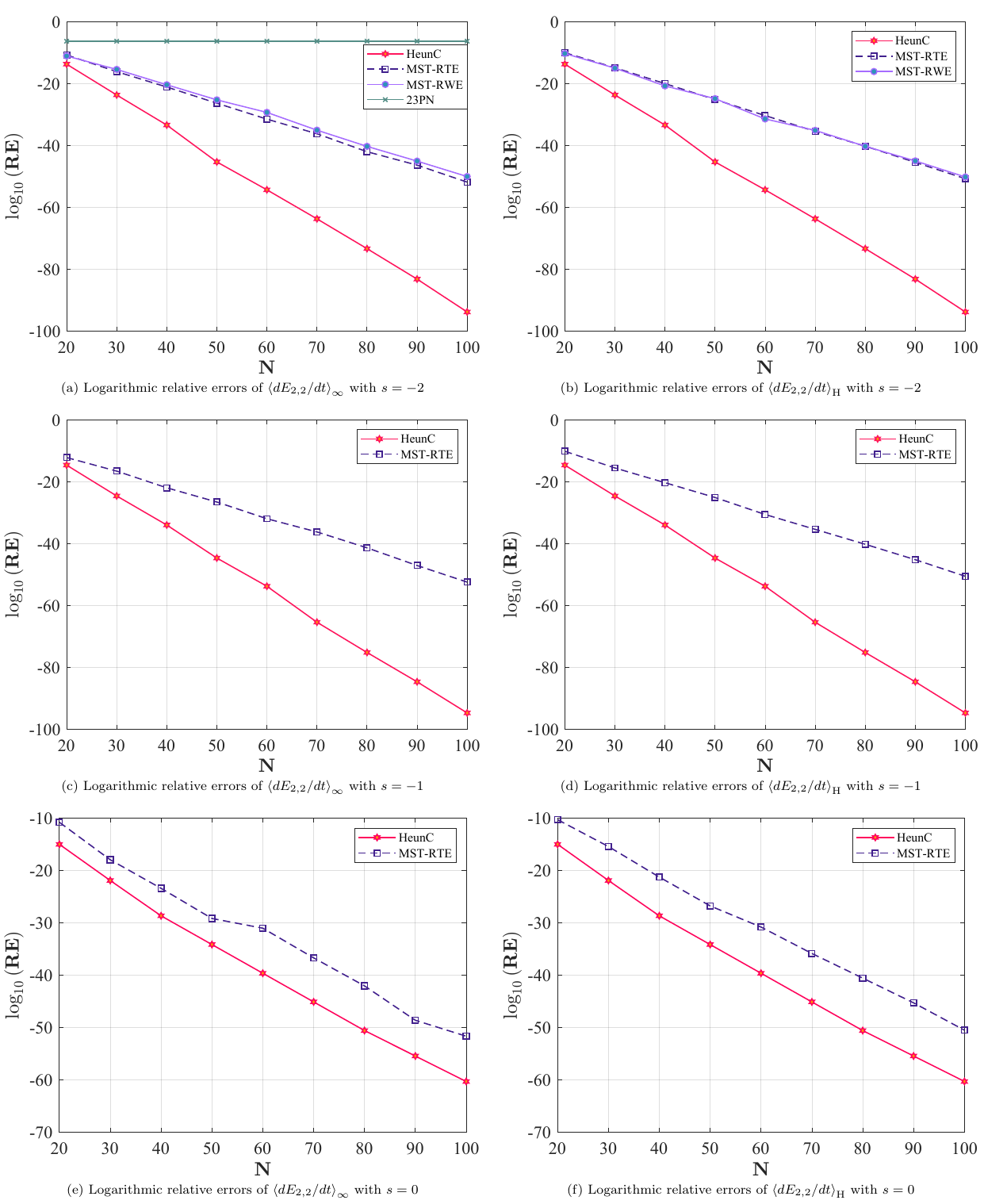}}
\caption{Logarithmic relative errors of energy fluxes $_s\left<{dE_{22}/dt}\right>_{\infty, {\rm H}}$ between four methods and the exact solution  with the  different floating-point numbers $\bf N$ at $r_{\rm ISCO}$. }\label{fig:FluxLogError}
\end{figure*}

%
\begin{table*}[htbp]
  \centering
  \caption{The comparison of four methods for energy fluxes $_s\left<{dE_{22}/dt}\right>_{\infty, {\rm H}}$ (angular momentum fluxes $_s\left<{dJ_{22}/dt}\right>_{\infty, {\rm H}}$) with the  different floating-point numbers $\bf N$ at $r_{\rm ISCO}$. In our tables, $_s\left<{dE_{22}/dt}\right>_{\infty, {\rm H}}$ is abbreviated as ${\dot E}_{\infty,{\rm H}}$.}\label{tab:flux-inf-hor}%
    \begin{threeparttable}
    \begin{tabular}{c|c|c|cccccccccc|c}
    \toprule
    $s$ & Flux & Method & $\bf N$ & 20   & 30   & 40   & 50   & 60   & 70   & 80   & 90   & 100  & Mean \\
    \midrule
    \multirow{15}[4]{*}{-2} & \multirow{8}[2]{*}{${\bf{RE}} ({\dot E}_\infty )$} & HeunC & $\bf  RE$ & 2.006E-14\tnote{*}\quad \quad& 2.168E-24 & 4.097E-34 & 4.845E-46 & 4.589E-55 & 1.882E-64 & 4.296E-74 & 6.043E-84 & 1.422E-94 & - \\
         &      &      & ROC  & -    & 9.966  & 9.724  & 11.927  & 9.024  & 9.387  & 9.642  & 9.852  & 10.628  & \textbf{10.019}  \\
         &      & MST-RTE & $\bf  RE$ & 1.892E-11 & 7.584E-17 & 8.281E-22 & 4.790E-27 & 3.400E-32 & 6.227E-37 & 9.808E-43 & 4.476E-47 & 1.422E-52 & -  \\
         &      &      & ROC  & -    & 5.397  & 4.962  & 5.238  & 5.149  & 4.737  & 5.803  & 4.341  & 5.498  & 5.141  \\
         &      & MST-RWE & $\bf  RE$ & 8.894E-12 & 4.043E-16 & 4.448E-21 & 5.755E-26 & 5.062E-30 & 9.023E-36 & 5.219E-41 & 8.896E-46 & 9.094E-51 & -  \\
         &      &      & ROC  & -    & 4.342  & 5.178  & 4.787  & 5.009  & 4.969  & 5.036  & 5.221  & 4.777  & 4.915  \\
         &      & 23PN & $\bf  RE$ & 5.417E-07 & 5.417E-07 & 5.417E-07 & 5.417E-07 & 5.417E-07 & 5.417E-07 & 5.417E-07 & 5.417E-07 & 5.417E-07 & -  \\
         &      &      & ROC  & -    & 0.000  & 0.000  & 0.000  & 0.000  & 0.000  & 0.000  & 0.000  & 0.000  & 0.000  \\
\cmidrule{2-14}         & \multirow{5}[2]{*}{${\bf{RE}} ({{{\dot E}_{\rm H}}} )$} & HeunC & $\bf  RE$ & 6.543E-11 & 1.300E-20 & 7.178E-31 & 4.771E-42 & 8.066E-52 & 5.255E-61 & 2.221E-70 & 1.545E-82 & 4.779E-92 & -  \\
         &      &      & ROC  & -    & 9.702  & 10.258  & 11.177  & 9.772  & 9.186  & 9.374  & 12.158  & 9.510  & \textbf{10.142}  \\
         &      & MST-RTE & $\bf  RE$ & 1.047E-10 & 1.440E-15 & 1.051E-20 & 7.309E-26 & 4.858E-31 & 3.720E-36 & 6.107E-41 & 3.496E-46 & 1.820E-51 & -  \\
         &      &      & ROC  & -    & 4.862  & 5.137  & 5.158  & 5.177  & 5.116  & 4.785  & 5.242  & 5.284  & 5.095  \\
         &      & MST-RWE & $\bf  RE$ & 5.510E-11 & 1.042E-15 & 1.877E-21 & 1.417E-25 & 3.886E-32 & 7.783E-36 & 5.495E-41 & 1.024E-45 & 6.537E-51 & -  \\
         &      &      & ROC  & -    & 4.723  & 4.959  & 4.888  & 5.056  & 3.749  & 6.238  & 4.768  & 4.990  & 4.874  \\ \midrule
    \multirow{11}[4]{*}{+2} & \multirow{5}[2]{*}{${\bf{RE}} ({\dot E}_\infty )$} & HeunC & $\bf  RE$ & 1.989E-14 & 2.385E-23 & 4.101E-34 & 2.988E-43 & 4.784E-55 & 1.884E-64 & 4.358E-74 & 6.028E-84 & 5.396E-94 & -  \\
         &      &      & ROC  & -    & 8.921  & 10.765  & 9.138  & 11.796  & 9.405  & 9.636  & 9.859  & 10.048  & \textbf{9.946}  \\
         &      & MST-RTE & $\bf  RE$ & 8.864E-12 & 3.031E-17 & 2.847E-22 & 1.452E-27 & 6.169E-32 & 2.483E-38 & 1.006E-41 & 9.288E-48 & 7.466E-54 & -  \\
         &      &      & ROC  & -    & 5.466  & 5.027  & 5.292  & 4.372  & 6.395  & 3.392  & 6.035  & 6.095  & 5.259  \\
         &      & MST-RWE & $\bf  RE$ & 2.195E-11 & 4.043E-16 & 4.448E-21 & 5.755E-26 & 5.062E-31 & 9.023E-35 & 5.219E-41 & 8.896E-46 & 9.094E-51 & -  \\
         &      &      & ROC  & -    & 4.735  & 4.959  & 4.888  & 5.056  & 3.749  & 6.238  & 4.768  & 4.990  & 4.923  \\
\cmidrule{2-14}         & \multirow{6}[2]{*}{${\bf{RE}} ({{{\dot E}_{\rm H}}} )$} & HeunC & $\bf  RE$ & 1.071E-10 & 2.911E-19 & 1.506E-30 & 2.264E-39 & 1.143E-50 & 2.823E-60 & 3.886E-70 & 2.468E-80 & 5.450E-91 & -  \\
         &      &      & ROC  & -    & 8.566  & 11.286  & 8.823  & 11.297  & 9.607  & 9.861  & 10.197  & 10.656  & \textbf{10.037}  \\
         &      & MST-RTE & $\bf  RE$ & 8.198E-11 & 1.137E-15 & 4.538E-21 & 8.956E-26 & 1.322E-30 & 1.855E-35 & 3.141E-41 & 6.077E-46 & 1.820E-51 & -  \\
         &      &      & ROC  & -    & 4.858  & 5.399  & 4.705  & 4.831  & 4.853  & 5.771  & 4.713  & 5.524  & 5.082  \\
         &      & MST-RWE & $\bf  RE$ & 1.096E-10 & 1.042E-15 & 1.877E-21 & 1.417E-25 & 3.886E-31 & 7.783E-36 & 5.495E-41 & 1.024E-45 & 6.537E-51 & -  \\
         &      &      & ROC  & -    & 5.022  & 5.744  & 4.122  & 5.562  & 4.698  & 5.151  & 4.730  & 5.195  & 5.028  \\
    \midrule
    \multirow{7}[4]{*}{$-1$} & \multirow{4}[2]{*}{${\bf{RE}} ({\dot E}_\infty )$} & HeunC & $\bf  RE$ & 2.533E-15 & 2.733E-25 & 1.084E-34 & 2.272E-45 & 1.694E-54 & 3.944E-66 & 6.398E-76 & 2.077E-85 & 1.775E-95 & -  \\
         &      &      & ROC  & -    & 9.967  & 9.402  & 10.679  & 9.127  & 11.633  & 9.790  & 9.489  & 10.068  & \textbf{10.019}  \\
         &      & MST-RTE & $\bf RE$ & 7.240E-13 & 2.935E-17 & 1.191E-22 & 3.002E-27 & 1.181E-32 & 6.267E-37 & 4.501E-42 & 8.646E-48 & 3.701E-53 & -  \\
         &      &      & ROC  & -    & 4.392  & 5.392  & 4.599  & 5.405  & 4.275  & 5.144  & 5.716  & 5.368  & 5.036  \\
\cmidrule{2-14}         & \multirow{4}[2]{*}{${\bf{RE}} ({{{\dot E}_{\rm H}}} )$} & HeunC & $\bf  RE$ & 4.540E-11 & 1.371E-20 & 1.127E-30 & 2.188E-41 & 4.866E-52 & 2.097E-61 & 1.684E-70 & 4.477E-80 & 6.692E-90 & -  \\
         &      &      & ROC  & -    & 9.520  & 10.085  & 10.712  & 10.653  & 9.366  & 9.095  & 9.575  & 9.825  & \textbf{9.854}  \\
         &      & MST-RTE & $\bf RE$ & 7.824E-11 & 3.102E-16 & 5.299E-21 & 8.606E-26 & 2.776E-31 & 4.241E-36 & 6.368E-41 & 6.556E-46 & 2.807E-51 & -  \\
         &      &      & ROC  & -    & 5.402  & 4.767  & 4.789  & 5.491  & 4.816  & 4.823  & 4.987  & 5.368  & 5.056  \\
    \midrule
    \multirow{7}[4]{*}{$+1$} & \multirow{4}[2]{*}{${\bf{RE}} ({\dot E}_\infty )$} & HeunC & $\bf  RE$ & 2.531E-15 & 2.696E-25 & 8.520E-35 & 1.049E-43 & 1.696E-54 & 3.947E-66 & 7.068E-76 & 2.085E-85 & 2.495E-95 & -  \\
         &      &      & ROC  & -    & 9.973  & 9.500  & 8.910  & 10.791  & 11.633  & 9.747  & 9.530  & 9.922  & \textbf{10.001}  \\
         &      & MST-RTE & $\bf RE$ & 2.822E-12 & 2.979E-17 & 2.804E-22 & 1.436E-26 & 4.333E-33 & 9.535E-38 & 7.576E-44 & 9.074E-48 & 1.175E-52 & -  \\
         &      &      & ROC  & -    & 4.976  & 5.026  & 4.291  & 6.520  & 4.657  & 6.100  & 3.922  & 4.888  & 5.048  \\
\cmidrule{2-14}         & \multirow{4}[2]{*}{${\bf{RE}} ({{{\dot E}_{\rm H}}} )$} & HeunC & $\bf  RE$ & 5.493E-11 & 1.285E-20 & 1.166E-30 & 2.105E-42 & 1.125E-50 & 4.089E-60 & 1.812E-70 & 4.657E-80 & 6.835E-90 & -  \\
         &      &      & ROC  & -    & 9.631  & 10.042  & 11.743  & 8.272  & 9.440  & 10.353  & 9.590  & 9.833  & \textbf{9.863}  \\
         &      & MST-RTE & $\bf RE$ & 4.628E-11 & 6.097E-17 & 4.777E-21 & 9.611E-26 & 3.575E-31 & 8.646E-36 & 8.646E-41 & 7.472E-46 & 1.068E-50 & -  \\
         &      &      & ROC  & -    & 5.880  & 4.106  & 4.696  & 5.429  & 4.616  & 5.000  & 5.063  & 4.845  & 4.955  \\
    \midrule
    \multirow{7}[4]{*}{$0$} & \multirow{4}[2]{*}{${\bf{RE}} ({\dot E}_\infty )$} & HeunC & $\bf  RE$ & 1.059E-15 & 1.296E-22 & 2.308E-29 & 7.493E-35 & 2.467E-40 & 8.204E-46 & 2.749E-51 & 3.757E-56 & 5.153E-61 & -  \\
         &      &      & ROC  & -    & 6.912  & 6.749  & 5.489  & 5.482  & 5.478  & 5.475  & 4.864  & 4.863  & \textbf{5.664}  \\
         &      & MST-RTE & $\bf RE$ & 1.708E-11 & 1.192E-18 & 4.098E-24 & 7.184E-30 & 9.991E-32 & 2.120E-37 & 9.476E-43 & 2.534E-49 & 2.300E-52 & -  \\
         &      &      & ROC  & -    & 7.156  & 5.464  & 5.756  & 1.857  & 5.673  & 5.350  & 6.573  & 3.042  & 5.109  \\ \cmidrule{2-14}
         & \multirow{4}[2]{*}{${\bf{RE}} ({{{\dot E}_{\rm H}}} )$} & HeunC & $\bf  RE$ & 4.250E-11 & 5.811E-20 & 3.617E-26 & 1.175E-31 & 3.867E-37 & 1.286E-42 & 4.310E-48 & 5.890E-53 & 8.077E-58 & -  \\
         &      &      & ROC  & -    & 8.864  & 6.206  & 5.488  & 5.483  & 5.478  & 5.475  & 4.864  & 4.863  & \textbf{5.840}  \\
         &      & MST-RTE & $\bf RE$ & 5.423E-11 & 4.007E-16 & 5.853E-22 & 1.812E-27 & 1.646E-31 & 1.479E-36 & 2.901E-41 & 5.421E-46 & 3.250E-51 & -  \\
         &      &      & ROC  & -    & 5.131  & 5.835  & 5.509  & 4.042  & 5.046  & 4.707  & 4.728  & 5.222  & 5.028  \\
    \bottomrule
    \end{tabular}%
    \begin{tablenotes}
        \footnotesize
        \item[*] The table data in this paper are abbreviated using scientific notation. For instance, $ 2.006\times10^{-14}$ is represented as  2.006E{-14}.

      \end{tablenotes}
\end{threeparttable}
\end{table*}

\begin{table*}[htbp]
  \centering
\caption{Relative errors between total energy fluxes $_s\left<{dE/dt}\right>_{\infty,{\rm H}}$ (total angular momentum fluxes $_s\left<{dJ/dt}\right>_{\infty,{\rm H}}$)  of four methods (${\bf N} =100$) and the exact solution at $r_{\rm ISCO}$(NA = not available). }\label{table:3spin-field-dEdt}
    \begin{tabular}{cccccccccc}
    \toprule
   &  &\multicolumn{4}{c}{Relative errors of $_s\left<{dE/dt}\right>_\infty$ }& \multicolumn{4}{c}{Relative errors of  $_s\left<{dE/dt}\right>_{\rm H}$ }  \\ \cmidrule(r){3-6} \cmidrule(r){7-10}
  {$s$} & Method & HeunC & MST-RTE & MST-RWE & 23PN & HeunC & MST-RTE & MST-RWE & 23PN \\
    \midrule
\multirow{2}{*}{$-2$} &  $\bf RE$   & 6.354E{-73}  &   2.769E{-52}  &  9.458E{-51} &  1.496E{-4}& 3.135E{-76}& 2.308E{-51} & 5.370E{-51}&  1.750E{-3}  \\
         & CPU(s)  &   197.9   &    591.7  &  588.1   &  NA   &   53.0    &   91.2   &  91.9    &  NA\\
    \midrule
\multirow{2}{*}{$+2$} & $\bf RE$   &  7.421E-72   &   7.993E-53   &  1.303E-50    &  NA        &  6.222E-74  & 9.458E-51   & 5.370E-51  & NA \\
                          & CPU(s)  &  193.6        &  571.3      &   576.1   &   NA            &  54.7    &   88.8  &   93.1   & NA \\
    \midrule
\multirow{2}{*}{$-1$} &  $\bf RE$   &  6.274E-72  &   1.136E-52  & NA &  NA   & 4.633E-76 & 5.302E{-52} & NA  & NA  \\
                           & CPU(s)  &    192.0    &    574.8      & NA &  NA   &    51.2   &  94.9      &  NA & NA  \\
\midrule
\multirow{2}{*}{$+1$} &  $\bf RE$   &  3.880E-72 &2.132E{-52}  & NA &  NA& 9.675E-75 & 9.867E{-51} & NA & NA  \\
                           & CPU(s)  &  189.5     &    599.2    &  NA&  NA&     51.4  &   98.0   &  NA    &  NA\\
\midrule
\multirow{2}{*}{$0$} &  $\bf RE$   &  5.539E-59  &    1.343E{-52}  & NA &  NA&  4.802E-53 &  3.451E{-51} & NA & NA  \\
                           & CPU(s)  &   139.2    &    295.5 &  NA   &  NA   &   29.2    &   52.5   &  NA    &  NA\\

    \bottomrule
    \end{tabular}%
\end{table*}

\begin{table*}[htbp]
  \centering
\caption{Relative errors  between energy fluxes $_s\left<{dE_{22}/dt}\right>_{\infty, {\rm H}}$ of  the four methods (${\bf N} =100$) and the exact solution in different orbital radius $r_0$. In our tables, $_s\left<{dE_{22}/dt}\right>_{\infty, {\rm H}}$ is abbreviated as ${\dot E}_{\infty,{\rm H}}$.}\label{table:Flux22}
    \begin{tabular}{ccccccccc}
    \toprule
    $s$    & Flux & $r_0$ & $5M$ & $10M$ & $20M$ & $50M$ & $100M$ & $200M$ \\
    \midrule
    \multirow{7}[4]{*}{$-2$} & \multirow{4}[2]{*}{${\bf{RE}} ({{{\dot E}_{\infty }}} )$} & HeunC & 1.597E-93 & 1.716E-97 & 8.967E-100 & 9.232E-100 & 1.809E-100 & 4.346E-100 \\
         &      & MST-RTE & 2.106E-52 & 2.002E-52 & 4.321E-52 & 4.066E-52 & 9.754E-54 & 1.015E-53 \\
         &      & MST-RWE & 2.123E-50 & 5.023E-52 & 2.120E-51 & 7.431E-52 & 1.593E-52 & 1.924E-52 \\
         &      & 23PN & 2.664E-05 & 7.473E-12 & 1.467E-18 & 1.512E-27 & 1.673E-34 & 1.629E-42 \\
\cmidrule{2-9}         & \multirow{3}[2]{*}{${\bf{RE}} ({{{\dot E}_{\rm H}}} )$} & HeunC & 1.029E-89 & 1.139E-91 & 4.150E-91 & 2.178E-88 & 1.688E-86 & 2.870E-85 \\
         &      & MST-RTE & 1.391E-50 & 2.807E-51 & 2.242E-51 & 1.287E-51 & 6.314E-53 & 1.248E-52 \\
         &      & MST-RWE & 1.307E-50 & 3.793E-51 & 2.724E-51 & 2.342E-51 & 8.606E-52 & 9.625E-52 \\
    \midrule
    \multirow{5}[4]{*}{$+2$} & \multirow{3}[2]{*}{${\bf{RE}} ({{{\dot E}_{\infty }}} )$} & HeunC & 1.413E-91 & 2.554E-97 & 1.601E-99 & 1.141E-100 & 4.792E-101 & 1.883E-100 \\
         &      & MST-RTE & 4.323E-53 & 2.095E-52 & 3.307E-52 & 2.149E-53 & 8.198E-54 & 2.720E-54 \\
         &      & MST-RWE & 2.123E-50 & 5.023E-52 & 2.120E-51 & 7.431E-52 & 1.593E-52 & 1.925E-52 \\
\cmidrule{2-9}         & \multirow{3}[2]{*}{${\bf{RE}} ({{{\dot E}_{\rm H}}} )$} & HeunC & 8.785E-90 & 9.953E-91 & 6.755E-91 & 5.800E-88 & 3.029E-85 & 9.663E-84 \\
         &      & MST-RTE & 1.282E-50 & 2.780E-51 & 7.089E-52 & 2.396E-52 & 1.798E-52 & 4.947E-53 \\
         &      & MST-RWE & 1.307E-50 & 3.793E-51 & 2.724E-51 & 2.342E-51 & 8.606E-52 & 9.625E-52 \\
    \midrule
    \multirow{3}[4]{*}{$-1$} & \multirow{2}[2]{*}{${\bf{RE}} ({{{\dot E}_{\infty }}} )$} & HeunC & 1.253E-93 & 1.080E-96 & 3.858E-99 & 3.665E-100 & 2.181E-98 & 1.043E-100 \\
         &      & MST-RTE & 5.168E-52 & 5.281E-54 & 1.161E-55 & 2.090E-53 & 2.500E-53 & 3.472E-54 \\
\cmidrule{2-9}         & \multirow{2}[2]{*}{${\bf{RE}} ({{{\dot E}_{\rm H}}} )$} & HeunC & 1.258E-89 & 1.325E-91 & 1.056E-90 & 5.104E-88 & 1.731E-86 & 5.447E-84 \\
         &      & MST-RTE & 6.124E-49 & 2.571E-51 & 1.711E-51 & 1.244E-52 & 3.509E-52 & 1.629E-53 \\
    \midrule
    \multirow{3}[4]{*}{$+1$} & \multirow{2}[2]{*}{${\bf{RE}} ({{{\dot E}_{\infty }}} )$} & HeunC & 1.248E-93 & 1.792E-97 & 1.361E-99 & 1.179E-101 & 2.179E-98 & 1.308E-100 \\
         &      & MST-RTE & 3.377E-53 & 2.792E-54 & 2.844E-52 & 1.730E-53 & 7.509E-53 & 2.153E-53 \\
\cmidrule{2-9}         & \multirow{2}[2]{*}{${\bf{RE}} ({{{\dot E}_{\rm H}}} )$} & HeunC & 2.158E-90 & 3.051E-91 & 1.517E-90 & 6.192E-88 & 4.198E-86 & 8.736E-84 \\
         &      & MST-RTE & 1.485E-50 & 1.066E-51 & 3.775E-51 & 3.217E-52 & 1.009E-52 & 1.195E-53 \\
    \midrule
    \multirow{3}[4]{*}{$0$} & \multirow{2}[2]{*}{${\bf{RE}} ({{{\dot E}_{\infty }}} )$} & HeunC & 2.622E-61 & 3.798E-61 & 6.767E-62 & 5.422E-63 & 7.733E-64 & 1.085E-64 \\
         &      & MST-RTE & 1.903E-53 & 1.857E-52 & 3.542E-53 & 4.881E-53 & 2.547E-54 & 3.736E-55 \\
\cmidrule{2-9}         & \multirow{2}[2]{*}{${\bf{RE}} ({{{\dot E}_{\rm H}}} )$} & HeunC & 1.034E-58 & 1.527E-56 & 1.134E-55 & 8.936E-55 & 3.812E-54 & 1.572E-53 \\
         &      & MST-RTE & 1.142E-50 & 1.898E-51 & 8.722E-52 & 2.350E-52 & 3.121E-53 & 1.668E-52 \\
    \bottomrule
    \end{tabular}%
  \label{tab:addlabel}%
\end{table*}

This section presents several numerical results to validate the adaptability of our method, and compare their accuracy with those which are already available in the literature for calculating scalar, electromagnetic, and gravitational radiations.
These methods in the literature include numerical integration (NI) method, high-order post-Newtonian expansion \cite{Fujita_2015} , MST-RTE method \cite{Mano_1996,Sasaki_2003} and MST-RWE method \cite{Mano1996RWE,Casals_2015}, whose codes are provided by the black hole perturbation toolkit (BHPT) \cite{BHPToolkit}.
In this subsection, there are abbreviations of some methods, MST-RTE and MST-RWE \footnote{MST-RWE first utilizes the MST method to solve the RW equation, and then uses the Chandrasekhar-Sasaki-Nakamura transformation to convert it into the solution of the homogeneous Teukolsky equation.}  are the MST methods for solving the radial Teukolsky equation (RTE) and Regge-Wheeler equation (RWE), respectively.
Meanwhile, 23PN represents 23th post-Newtonian order expansion, and the HeunC method is the method proposed in this paper.
The BHPT is a collection of multiple scattered black hole perturbation theory codes, which have been developed by various individuals or groups over several decades.
Although the BHPToolkit yields a formal solution of the confluent Heun function that solves the homogenous Teukolsky equation, it solely provides  the ingoing wave solution $R_{\ell m\omega }^{{\rm{in }}}$, but the outgoing wave solution $R_{\ell m\omega }^{{\rm{up}}}$has not been given so far. Therefore, this formal solution
(This solution is composed of the confluent Heun function, but it is different from our solution.) of the BHPToolkit is an imperfect method and cannot calculate radiation fluxes.

In numerical calculations for partial differential equations, the rate of convergence (ROC) is a measure of how well a numerical solution approaches the exact solution as the spatial or temporal resolution is increased. It provides insight into the accuracy and reliability of the method.
Therefore, we introduce the ROC of the floating-point numbers\footnote{$\bf N$ is the software floating-point numbers, which can affect the computational efficiency. The larger $\bf N$, the more calculation time is required.} $\bf N$ to evaluate the speed of convergence for the computational methods.
And the ROC of floating-point numbers is defined as \cite{Li2008,CHEN2020125009,Chen2020b}
\begin{equation}
  {\rm{ROC}} = {\log _{10}}\left( \frac{{\bf{RE}}_{\bf N}}{{\bf{RE}}_{{\bf N}+10}} \right),
\end{equation}
where ${{\bf{RE}}_{{\bf N}}}$ and ${{\bf{RE}}_{{\bf N}+10}}$ are the relative error (RE) of ${\bf N}$ and ${\bf N}+10$, respectively.
The exact solution in this paper is the results of the MST-RTE method with ${\bf N} =200$. In our Tables, CPU is computing time. All the numerical experiments are conducted on an Intel Core i7-12700H 2.70 GHz processor.

To validate the correctness and applicability of the HeunC method, we test two kind of radiation fluxes $_s\left<{dE_{\ell m}/dt}\right>_{\infty, {\rm H}}$ propagates along the ($+r$) direction.
\Cref{fig:FluxInf,fig:FluxHor} show logarithmic values of $_s\left<{dE_{\ell m}/dt}\right>_{\infty, {\rm H}}$ of the three perturbation fields (scalar, electromagnetic, and gravitational perturbations) with different modes, over the domain $[5M,15M] $.
In order to see the value of the energy fluxes at $r_{\rm{ISCO}}$\footnote{The innermost stable circular orbit(ISCO) is only defined in the equatorial plane, that is $r_{\rm ISCO}=6M$.}, we draw a slice diagram of the three perturbation fields as shown in \Cref{fig:Flux2Inf,fig:Flux2Hor}.
It can be seen from \Cref{fig:FluxInf,fig:FluxHor,fig:Flux2Inf,fig:Flux2Hor}  that when ${\bf N} =50$,  the energy fluxes obtained by the HeunC method exhibit a commendable agreement with the results obtained by the MST-RTE method.
This numerical simulation confirms the physical phenomenon that the closer a point particle approaches a black hole, the greater the amount of radiative energy flux it generates. Furthermore, when $r_0$ is fixed, the maximum energy flux across all modes are $_s\left<{dE_{22}/dt}\right>_{\infty, {\rm H}}$.
According to the properties of the spin-weighted spherical harmonic function ${}_{s}Y_{\ell m\omega }$, when $|\ell|-|m|={\rm odd}$, ${}_{0}Y_{\ell m\omega }=0$. From \Cref{eq:dEdtInfS0,eq:dEdtHorS0}, when ${}_{0}Y_{\ell m\omega }$ is equal to zero, the energy fluxes are also zero.
Therefore, when $| \ell| - |m| = {\rm odd} $, the logarithmic energy flux of the scalar perturbation field tends to infinity.
This property can be verified by the logarithmic energy fluxes of the scalar perturbation field in  \Cref{fig:Flux2Inf,fig:Flux2Hor}.

To facilitate a comprehensive comparison of the accuracy and efficiency of the HeunC method with other existing methods, we provide numerical comparisons of energy fluxes $_s\left<{dE_{22}/dt}\right>_{\infty, {\rm H}}$ of three perturbation fields obtained from four methods for different floating-point numbers ${\bf N}$ at $r_{\rm ISCO}$ in \Cref{tab:flux-inf-hor}. Meanwhile, the error comparison of total radiation fluxes is shown in \Cref{table:3spin-field-dEdt}.
Because the relative errors in \Cref{tab:flux-inf-hor} and \Cref{table:3spin-field-dEdt} are not less than $10 ^ {\bf N}$, the error results of the energy flux are the same as those of the angular momentum flux (their relationship can be seen in \cref{eq:dJdt}).
Analyzing the ROC presented in \Cref{tab:flux-inf-hor}, it can be concluded that the HeunC method is a 10th-order method for the calculation of electromagnetic and gravitational luminosities, and the MST-RTE and MST-RWE methods are 5th-order methods.
Despite a decrease in ROC as ${\bf N}$ increases for the scalar perturbation field, the HeunC method maintains superior accuracy in calculating the energy fluxes compared to the MST-RTE method.
This behavior is evident from the observations presented in \Cref{fig:FluxLogError}. As $\bf N$ increases, the HeunC method exhibits a notably larger decrease in relative errors in contrast to both the MST-RTE and MST-RWE methods. These results show that the convergence speed of HeunC is faster than that of the MST-RTE and MST-RWE methods.

To explore the precision of energy fluxes across various modes, \Cref{table:3spin-field-dEdt} presents the relative errors of total energy flux for three  perturbation fields. To obtain the total energy flux, it is necessary to sum $\ell$ in the range $2\leq \ell \leq 25$ ($2\leq \ell \leq 11$) for the infinity (horizon) part. It can be seen from \Cref{table:3spin-field-dEdt} that the HeunC method calculates electromagnetic and gravitational energy fluxes, and obtains more accurate and efficient results than other methods. Even when calculating the scalar energy fluxes, the accuracy is attenuated, but it is still superior to other methods.
\Cref{table:Flux22} exhibits the relative errors of energy fluxes within a region $r_0\in [5M,200M] $, aiming to assess the precision of four distinct methods as $r_0$ is close to the horizon or infinity. Notably, in both near the horizon and at infinity, the HeunC method showcases exceptional accuracy in calculating the energy fluxes, surpassing the other three methods by a significant margin.
Moreover, ROC of 23PN method presented in \Cref{tab:flux-inf-hor} and relative errors in \Cref{table:Flux22}, reveal that  the relative error of the 23PN method decreases as $r_0$ increases, while remaining unaffected by the floating-point numbers.
This also exposes a limitation of the 23 PN method: it exhibits inaccuracies in the vicinity of the horizon,  but when $r_0\rightarrow \infty $, the results obtained from this method become increasingly accurate.

All numerical results of this test show that the computational accuracy and efficiency of the HeunC method surpasses that of the other three methods to a significant extent. This is advantageous for waveform construction.

\section{Conclusion}\label{sec:Conclusion}
The purpose of this paper is to present an exact method to calculate gravitational, electromagnetic and scalar radiation fluxes for any type-D BHs.
We first reformulate the radial Teukolsky equations into a general form, which is classified into multiple types based on $\Delta_n$.
Then, we focus our attention on solving the $\Delta_2$-type Teukolsky equation which includes ten kinds of Teukolsky equations shown in \Cref{table:GFTE}.
The general solution \eqref{eq:D2GSol1} of homogenous form \eqref{eq:HRTEs} of  the $\Delta_2$-type Teukolsky equation is constituted by the linear combination of two linearly independent solutions which are expressed by confluent Heun functions.
This solution, benefiting from the analytical asymptotic expression \eqref{eq:87} of the confluent Heun function at infinity which is obtained in this paper for the first time, can be applied to the physical models corresponding to various boundary conditions.
Meanwhile, ingoing wave and outgoing wave solutions $R_{\ell m\omega }^{{\rm{in,up}}}$ can directly be obtained from the exact general solution according to boundary conditions.
In a significant contribution, , we provide a complete and strict derivation process for expressing $R_{\ell m\omega }^{{\rm{in,up}}}$ in terms of confluent Heun functions.
Consequently, the exact solution of inhomogeneous GTEs is found utilizing Green's function.
It's worth noting that these exact solutions are not subject to any constraints (such as slow-motion, low-frequency, and weak-field).
Next, Green's function method is employed to evaluate $Z^{\infty,{\rm H}}_{\ell m\omega}$, which is used in \Cref{eq:dEdtInf,eq:dEdtH} to compute the energy flux of the gravitational, electromagnetic and scalar waves.
Finally, the HeunC Method is used for numerical simulations of the radiation fluxes of the Schwarzschild black hole, and the results are entirely satisfactory in comparison with the analytical (23th post-Newtonian order expansion) and numerical methods \cite{Fujita_2004,Fujita_2005}. The findings are summarized in detail as follows:

1. For calculating the gravitational, electromagnetic, and scalar fluxes for Schwarzschild black hole, our method can accurately calculate any mode of the entire spatial region $r_0\in[r_+,\infty)$, and its accuracy is much higher than the results obtained by other methods. Additionally, the PN expansion suffers from slow convergence and limitations in the low-frequency approximation, making its results near the event horizon less precise. Even in the far distance, our approach produces superior results to those obtained using the PN expansion, MST-RTE, and MST-RWE methods. And the computational time of our method is less than half of other numerical methods. Thus, our method is superior to both the MST and PN methods in terms of its exceptional precision and efficiency.

2. Fujita and Tagoshi have demonstrated that the MST method converges very fast \cite{Fujita_2004,Fujita_2005}. However, they did not provide a mathematical evaluation of the convergence specifically for floating-point numbers. The increase in the floating-point numbers is accompanied by improved accuracy but also increased computation time. To this end, we introduce ROC for floating-point numbers.
By comparing the ROC of the four methods, it has been determined that HeunC is a 10th-order method, while MST is 5th-order methods.
But, the 23PN method is not convergent about the floating-point numbers.
These comparisons mean that for each additional floating-point number $\bf N$, the relative error of the HeunC method will be multiplied by $10^{-1}$, whereas the relative error of the MST and NI methods will be multiplied by $5\times 10^{-1}$.

3. Due to the complexity of the source term of the neutron star \cite{osti_4023275}, the radiation fluxes calculated in this paper only considers the gravitational, electromagnetic, and scalar perturbations. However, if the source term of the neutron star is known, our method can also find the radiation fluxes of the neutron star. In theory, our method can be applied to compute the radiation fluxes of any perturbation fields, as long as their source terms are known.

All in all, our method exhibits high precision, high efficiency, and excellent applicability. It can be applied to calculate the radiation fluxes of any perturbation fields quickly and accurately.
Accurate and efficient energy flux calculation is the basis for obtaining GW waveforms.
Therefore, in future work, we will use the HeunC method to derive the radiation fluxes and GW waveforms for the test particles moving around a rotating black hole in a general orbit.
Furthermore, our general solution is beneficial for solving the Teukolsky equation in complicated spacetimes ( Kerr-Newman anti-de Sitter BHs \cite{Khanal_1983,Suzuki_1998,Suzuki_1999} ), modified gravity theories \cite{Li_2023}, or effective-one-body theories based on post-Minkowskian approximation \cite{Jing_2022,Jing_2023}.

\section*{Acknowledgement}
This work was supported by the Grant of NSFC No. 12035005,
and National Key Research and Development Program of China No. 2020YFC2201400.
\appendix

\section{Expression of ${}_s \tilde Z_{\ell m\omega}^{{\rm H},\infty}$, ${_s{\beta _{\ell m\omega }}}$, and ${{}_s{\alpha _{\ell m\omega }}}$}\label{app:expression}

Substituting the point sources ${}_s T_{\ell m\omega}$ of gravitational, electromagnetic, and scalar perturbations in Ref. \cite{Teukolsky1973}   into \cref{eq:Z0} respectively, we can obtain the expression of ${}_s \tilde Z_{\ell m\omega}^{{\rm H},\infty}$.

\begin{equation}\label{eq:Zint}
  {}_s \tilde Z_{\ell m\omega}^{{\rm H},\infty} = \frac{1}{{W_{\rm C} }}\int_{ - \infty }^\infty  d t{e^{i\omega t - im\varphi (t)}}({}_s {\cal I}_{\ell m\omega }^{\infty,{\rm H}})_{r = r(t),\theta  = \theta (t)},
\end{equation}

For the gravitational, electromagnetic, and scalar perturbations, we can incorporate the general formula \eqref{eq:Zint}. Each perturbation field introduces distinct functions ${}_s {\cal I}_{\ell m\omega }^{\infty,{\rm H}}$. In this section, we provide the specific form of ${}_s {\cal I}_{\ell m\omega }^{\infty,{\rm H}}$ associated with different perturbation fields.

\subsection{Gravitational field}

For the general formula \eqref{eq:Zint}, ${}_s {\cal I}_{\ell m\omega }^{\infty,{\rm H}}$ of the gravitational perturbation field of the spin-weight $s = -2$, and the coefficients in fluxes \eqref{eq:dEdtInf} and \eqref{eq:dEdtH} can be given by

\begin{align}
{}_{-2}{\cal I}_{\ell m\omega }^{\infty,\rm{H}} &= \frac{8 \pi}{\Upsilon_t }\left[ {\left( {{A_{nn0}} + {A_{\bar mn0}}+{A_{\bar m\bar m0}}}\right)R_{\ell m\omega }^{{\rm{in,up}}}} \right. \nonumber \\
 &- \left( {{A_{\bar mn1}} + {A_{\bar m\bar m1}}} \right){\left( {R_{\ell m\omega }^{{\rm{in,up}}}} \right)^\prime }{\left. + {A_{\bar m\bar m2}}{{\left( {R_{\ell m\omega }^{{\rm{in,up}}}} \right)}^{\prime \prime }} \right]},
\end{align}
and
\begin{subequations}\label{}
\begin{align}\label{}
{}_{-2}\alpha_{\ell m\omega } &= \frac{{256{{(2M{r_{\rm{H}}})}^5}({\omega ^2} + 4{{\tilde \epsilon }^2})({\omega ^2} + 16{{\tilde \epsilon }^2}){\omega ^4}}}{{|{{\bf{C}}_{ - 2}}{|^2}}},\\
{}_{-2}\beta_{\ell m\omega }& =1.
\end{align}
\end{subequations}
where $\tilde \epsilon  = {(4{r_{\rm{H}}})^{ - 1}}$ and ${{|{\bf{C}}_{-2}|{^2}}} = {\lambda ^2}{(\lambda  + 2)^2}  + 144 {\omega ^2} {M^2}.$ And the explicit form of $A_{nn0}$ and other terms can be found and derived in Ref. \cite{Sasaki_2003,Teukolsky1973}.

Similarly,  ${}_s {\cal I}_{\ell m\omega }^{\infty,{\rm H}}$ of the spin-weight $s=+2$  and the coefficients in fluxes \eqref{eq:dEdtInf} and \eqref{eq:dEdtH} case can be given by
\begin{align}
{}_{+2}{\cal I}_{\ell m\omega }^{\infty,\rm{H}} &= \frac{8 \pi}{\Upsilon_t } \left[ {\left( {{A_{ll0}} + {A_{lm0}}+{A_{ mm0}}}\right)R_{\ell m\omega }^{{\rm{in,up}}}} \right. \nonumber \\
 &- \left( {{A_{lm1}} + {A_{mm1}}} \right){\left( {R_{\ell m\omega }^{{\rm{in,up}}}} \right)^\prime } {\left. + {A_{mm2}}{{\left( {R_{\ell m\omega }^{{\rm{in,up}}}} \right)}^{\prime \prime }} \right]},
\end{align}
and
\begin{subequations}\label{}
\begin{align}\label{}
{}_{+2}\alpha_{\ell m\omega} &= \frac{{{\omega ^2}}}{{128r_{\rm{H}}^3\left( {{\omega ^2} + 4{\epsilon ^2}} \right)}} ,\\
{}_{+2}\beta_{\ell m\omega}& = \frac{{16{\omega ^8}}}{{|{{\bf{C}}_2}{|^2}}}.
\end{align}
\end{subequations}
with $|{{\bf C}_2}|^2 = {(\lambda  + 4)^2}{(\lambda  + 6)^2} + 144{M^2}{\omega ^2}$.

\subsection{Electromagnetic field}
For the general formula \eqref{eq:Zint}, ${}_s {\cal I}_{\ell m\omega }^{\infty,{\rm H}}$ of the electromagnetic perturbation field of the spin-weight $s=-1$, and the coefficients in fluxes \eqref{eq:dEdtInf} and \eqref{eq:dEdtH} can be given by
\begin{equation}\label{}
  {\cal I}_{\ell m\omega }^{\infty,\rm{H},-1} =  {\frac{\mathfrak{A}_{-}}{\Delta }{R_{\ell m\omega }^{{\rm{in,up}}}} - \frac{ \mathfrak{B}_{-}}{\Delta }{\left( {R_{\ell m\omega }^{{\rm{in,up}}}} \right)^\prime } } ,
\end{equation}
and
\begin{subequations}\label{}
\begin{align}\label{}
{}_{-1}{\alpha _{\ell m\omega }} &= \frac{{64{r_{\rm{H}}}{M^3}{\omega ^4}\left( {4{r_{\rm{H}}}^2{\omega ^2} + 1} \right)}}{{{\lambda ^2}}},\\
{}_{-1}{\beta _{\ell m\omega }}&  =2.
\end{align}
\end{subequations}
where the explicit form of ${\mathfrak{A}_{\pm}}$ and ${\mathfrak{B}_{\pm}}$ can be found in Eq. (48) of Ref. \cite{Torres_2022}

Similarly,  ${}_s {\cal I}_{\ell m\omega }^{\infty,{\rm H}}$ of the spin-weight $s=+1$  and the coefficients in fluxes \eqref{eq:dEdtInf} and \eqref{eq:dEdtH} case can be given by
\begin{equation}\label{}
  {}_{+1}{\cal I}_{\ell m\omega }^{\infty,\rm{H}} =  {{\mathfrak{A}_{+}} R_{\ell m\omega }^{{\rm{in}},{\rm{up}}} - \frac{ \mathfrak{B}_{+}}{\Delta }{\left( \Delta R_{\ell m\omega }^{{\rm{in}},{\rm{up}}}  \right)^\prime } } ,
\end{equation}
and
\begin{subequations}\label{}
\begin{align}\label{}
{}_{+1}{\alpha _{\ell m\omega }} &=\frac{{{\omega ^2}}}{{4r_{\rm H}}},\\
{}_{+1}{\beta _{\ell m\omega }}& =\frac{{8{\omega ^4}}}{{{{(\lambda  + 2)}^2}}},
\end{align}
\end{subequations}

\subsection{Scalar field}
For the general formula \eqref{eq:Zint}, ${}_s {\cal I}_{\ell m\omega }^{\infty,{\rm H}}$ of the scalar perturbation field of the spin-weight $s=0$, and the coefficients in fluxes \eqref{eq:dEdtInf} and \eqref{eq:dEdtH} can be given by
\begin{equation}\label{eq:scala-I}
{}_{0} { {\cal I}_{\ell m\omega }^{\infty,\rm{H}} }= \frac{ -4 \pi  {r^2_0} }{\Upsilon_t}{}_{0}Y_{\ell m\omega } R_{\ell m\omega }^{{\rm{in}},{\rm{up}}},
\end{equation}
and
\begin{subequations}\label{}
\begin{align}\label{}
{}_{0}{\alpha _{\ell m\omega }} &= 4M{\omega ^4},\\
{}_{0}{\beta _{\ell m\omega }}& =1.
\end{align}
\end{subequations}
where ${}_{s}Y_{\ell m\omega }$ is the spin-weighted spherical harmonics.

\bibliography{mybibfile}

\end{document}